\documentclass[a4paper,11pt]{article} 
\pdfoutput=1 

\usepackage{jcappub} 
\usepackage[utf8]{inputenc} 

\usepackage{epstopdf}
\usepackage{subcaption} 
\usepackage{amsmath}
\usepackage{amsfonts}

\newcommand{\BigO}[1]{\ensuremath{\operatorname{\mathcal{O}}\left(#1\right)}}

\makeatletter
\gdef\@fpheader{}
\g@addto@macro\bfseries{\boldmath}
\makeatother

\title{(P)reheating and gravitational waves in $\alpha$-attractor models}

\author{Tomasz Krajewski,}
\author{Krzysztof Turzy\'nski}

\affiliation{Institute of Theoretical Physics, Faculty of Physics, University of Warsaw, \\ Pasteura 5, 02-093 Warsaw, Poland}

\emailAdd{tomasz.krajewski@fuw.edu.pl}
\emailAdd{krzysztof.turzynski@fuw.edu.pl}

\arxivnumber{2204.12909}

\abstract{
We study post-inflationary evolution in $\alpha$-attractor T-models of inflation. We consider the dynamics of both scalar fields present 
in these models: the inflaton and the spectator, as a negative field-space curvature may lead to geometrical destabilization of the spectator.
We perform state-of-the-art lattice simulations
with a dedicated numerical code optimized for those models.
We corroborate earlier findings that the perturbations of the spectator field are much more unstable than the perturbations of the inflaton field,
so the dynamics of early stages of preheating is dominated by the evolution of spectator perturbations.
We also calculate the spectrum of gravitational waves
originating from scalar fluctuations 
in order to determine if
the $\alpha$-attractor T-models can
be constrained or even ruled out by present cosmological observations, but not by direct searches of gravitational waves. 
}

\keywords{inflation, preheating, lattice simulations}

\begin{document}
\maketitle
\flushbottom

\section{Introduction}

By now, cosmological inflation has become a firmly rooted part of the standard cosmological model 
(see e.g.\ \cite{Mukhanov:2005} for a pedagogical introduction). 
While the increasing accuracy of the determination of
cosmological parameters has significantly narrowed down
the range of empirically allowed inflationary models,
there are still a number of theoretically motivated possibilities.
In addition, there are also many ideas about
the \emph{reheating}, i.e.\ the mechanism of
populating the Universe with degrees of freedom collectively
satisfying the equation of state characteristic of radiation.
Although  there exist possible scenarios for reheating, well embedded in the general framework of quantum field theory \cite{Traschen:1990sw,Kofman:1994rk}, there are typically no observables that could 
at present distinguish between various proposed scenarios.    

One of the actively studied mechanisms of reheating is the so-called \emph{preheating}.
It relies on the fact that the fluctuations of the inflaton and other scalar fields -- in the background of the homogeneous inflaton oscillating around the minimum of its potential -- can be unstable, which potentially leads to mode amplification and particle production \textit{via} a  parametric resonance \cite{Kofman:1994rk,Shtanov:1994ce,Kofman:1997yn}. 

A notable example of a class of models that are naturally equipped with the preheating mechanism is the \mbox{T-model} subset of $\alpha$-attractor models of inflation \cite{Carrasco:2015rva}, which continue to attract a lot of theoretical interest.
It is known that in a single-field effective theory of the inflaton field stemming from these models,  
the inflaton experiences self-resonance \cite{Amin:2011hj} at the end of inflation and its perturbations may become highly unstable. Once they dominate the Universe, the effective equation of state quickly approach that of radiation \cite{Lozanov:2016hid,Lozanov:2017hjm}.
As a consequence, the radiation-dominated era effectively begins very soon after inflation,
which greatly reduces the theoretical uncertainty
stemming from interpreting cosmological observations
in the context of our ignorance about the reheating era \cite{Liddle:2003as,Adshead:2010mc,Creminelli:2014oaa,Dai:2014jja,Martin:2014nya,Cook:2015vqa,Ueno:2016dim,Eshaghi:2016kne}.

However, even in the minimal supergravity construction, one should in principle consider both real degrees of freedom present in the scalar part of the chiral multiplet. 
It turns out that there are interesting regions of the parameter space of $\alpha$-attractor models which cannot be described by a single-field effective theory. 
The second scalar field, residing in the chiral multiplet of
the supergravity construction typicaly plays the role of a passive
\emph{spectator}.
It is heavy when the 
modes that we now observe through temperature anisotropies of
the cosmic microwave background (CMB) radiation
leave the Hubble radius; its presence can, therefore, be safely neglected for the calculation of the power spectrum of the curvature perturbations. 
This assumption breaks down at  the end of inflation, when the spectator field becomes transiently tachyonic and unstable, so its perturbations may eventually dominate the Universe.
This is possible because the non-canonical form of the kinetic part of the Lagrangian gives rise to geometrical destabilization \cite{Renaux-Petel:2015mga}.

It is therefore interesting to investigate numerically the impact of the spectator field on the dynamics of reheating, using the full
two-field approach. In the past,
we made the first step in this direction \cite{Krajewski:2018moi},
firmly demonstrating that the spectator inhomogeneities quickly dominate
over those of the inflaton. As a consequence, we showed that
preheating lasts even shorter than previously thought. Moreover,  
due to nonlinear interactions, the dramatically fast increase of
the amplitude of spectator fluctuations is translated to inflaton
perturbations and the entire system is destabilized. However, we have not been able
to resolve the instability in sufficient detail to track the effective equation of state of the Universe, because of the limited lattice size. 

In the present paper, we attempt to address this shortcoming of 
our previous analysis by increasing the lattice size. 
We also study the spectrum of gravitational waves that
are produced during preheating out of scalar fluctuations.
We find that the instability is so strong that increasing the linear
lattice size by a factor of 4 (and thus increasing the lattice volume
by a factor of 64) compared to our previous study does not allow 
for resolving all aspects of the 
dynamics of the unstable modes. We thereby confirm
that the instability of the spectator field is a crucial element
of the evolution of the Universe and that the preheating occurs
almost instantaneously. We also analyze the spectrum of the gravitational
waves produced out of the scalar perturbations and find that the amplitude
of the gravitational waves grows significantly.

Our paper is organized as follows. In Section~\ref{sec:Tmodel}, 
we recapitulate the main ingredients of the $\alpha$-attractor T-models
that we consider. In Section~\ref{sec:three},
we describe our methods of lattice simulations for scalar degrees
of freedom and present the results of those simulations.
In Section~\ref{sec:GWs_spectrum}, we present our results for 
the spectrum of gravitational waves. Technical remarks and
supplementary results which help understand our methods and results,
but which are not necessary for inferring the main message of our paper
are relegated to appendices.

\section{$\alpha$-attractor T-models of inflation}
\label{sec:Tmodel}

\subsection{Presentation of the model\label{sec:presentation_of_model}}
In this paper, we will consider the simplest and most thoroughly studied class of $\alpha$-attractor models of inflation, the so-called  T-models. They can be naturally implemented in the supergravity with hyperbolic geometry as follows \cite{Ferrara:2013rsa,Kallosh:2013maa,Kallosh:2013daa,Kallosh:2013yoa,Kallosh:2015lwa,Carrasco:2015rva,Carrasco:2015uma,Ferrara:2016fwe,Achucarro:2017ing}.
In so-called half-plane variables, $T$~and~$\bar{T}$, the K\"ahler potential takes the form
\begin{equation}
K_H=-\frac{3\alpha}{2}\log{\left(\frac{(T-\bar{T})^2}{4T\bar{T}}\right)}+S\bar{S}\label{eq:kahler_potential}
\end{equation}
with parameter $\alpha > 0$ and the superpotential is
\begin{equation}
W_H=\sqrt{\alpha}\mu S{\left(\frac{T-1}{T+1}\right)^n}, \label{eq:super_potential}
\end{equation}
where $n > 0$ and $\mu$ is a constant parameter.

As shown in \cite{Carrasco:2015uma}, the superfield $S$ can be stabilized during and after inflation. Therefore, we will neglect its contribution into evolution of the Universe and set $S = 0$. Then, the scalar sector of the model can be expressed in terms of two real scalar fields $\phi$ and $\chi$, which are related to the scalar component of the superfield $T$ by
\begin{align}
\left|\frac{T-1}{T+1}\right|^2 &= \left(\frac{\cosh(\beta\phi)\cosh(\beta\chi) - 1}{\cosh(\beta\phi)\cosh(\beta\chi) + 1}\right), && \textrm{where} & \beta &:= \sqrt{\frac{2}{3\alpha}}.
\end{align}
This choice is beneficial from the point of view of lattice simulations since it leads to a~particularly simple form of the field-space metric in the kinetic term of the scalar Lagrangian: 
\begin{equation}
\mathcal{L}=-\frac{1}{2}\Big(\partial_\mu\chi\partial^\mu\chi+\textrm{e}^{2b(\chi)}\partial_\mu\phi\partial^\mu\phi\Big)-V(\phi,\chi). \label{eq:field_space_metric}
\end{equation}
where $b(\chi)\equiv\log(\cosh(\beta\chi))$. The potential of the model in this variables reads 
\begin{equation}
V(\phi,\chi)=M^4\left(\frac{\cosh(\beta\phi)\cosh(\beta\chi) - 1}{\cosh(\beta\phi)\cosh(\beta\chi) + 1}\right)^n{\left(\cosh(\beta\chi)\right)^{2/\beta^2}}, \label{eq:potential}
\end{equation}
with $M^4=\alpha\mu^2$.

With such a field-space metric, the reparametrization invariant field-space curvature can be easily calculated
\begin{equation}
\mathbb{R}=-2(b'^2+b'')=-2\beta^2=-\frac{4}{3\alpha}, \label{eq:field_space_curvature}
\end{equation}
where $'$ denotes the derivative of the function with respect to its argument. We note that $\mathbb{R}$ is nonzero, thus one cannot canonically normalize both degrees of freedom at once. Moreover, $\mathbb{R}$ is negative, which is crucial for the possibility of geometrical destabilization. Finally, the value of $\mathbb{R}$ is determined by $\alpha$, and the inverse proportionality means that the size of $\mathbb{R}$ increases with decreasing value of $\alpha$ parameter, thus one expect the stronger destabilization for smaller values of $\alpha$.

\subsection{Inflationary trajectory}
It was shown in \cite{Carrasco:2015rva} that T-models presented in Section~\ref{sec:presentation_of_model} admit an inflating solution with the inflationary trajectory proceeding along $\chi=0$. Moreover, in \cite{Iarygina:2018kee} the case $\chi \neq 0$ was investigated and it 
was shown that there are
two stages of inflation; the first one proceeds nearly 
along $\phi = const$ and the second one with $\chi$ being exponentially close to $0$. 
However, for reasonable values of the ratio $\frac{\chi_0}{\phi_0}$ of 
initial values of the field strengths, 
the number of e-folds of the second stage is much larger than that of the first stage. Therefore, the generic observational prediction of these models for the CMB anistotropies is that of the usual single-field inflation. Remarkably, the model is consistent with Planck data for a wide range of its parameters (see e.g. \cite{Kallosh:2013yoa,Kallosh:2021mnu}).

The inflationary trajectory can be effectively described in terms of the canonically normalized inflaton field $\phi$ with the potential 
\begin{equation}
V(\phi, 0)=M^4\tanh^{2n}\left(\frac{\beta|\phi|}{2}\right).
\end{equation}
Setting $\chi=0$ during inflation, the trajectory is the solution of following equations of motion for the space homogeneous quantities $H := \dot{a}/a$ and $\phi$:
\begin{align}
H^2 &= \frac{1}{3}\left(\frac{1}{2}\dot{\phi}^2+V(\phi,0)\right), & \ddot{\phi}+3H\dot{\phi}+V_\phi(\phi,0) &= 0, \label{eq:background_eom}
\end{align}
where $V_\phi$ stands for a derivative in the $\phi$ direction ($V_\phi\equiv \frac{\partial V}{\partial \phi}$) and $\dot{}$ denotes the derivative with respect to cosmic time $t$. We introduced $a(t)$ for the scale factor of Friedmann-Robertson-Walker metric.

As usual in single-field models, the inflaton accelerates during inflation and eventually leaves the slow-roll regime defined as $\epsilon_H\equiv-\dot{H}/H^2 \ll 1$ and $\eta_H\equiv-\ddot{H}/(2\dot{H} H) \ll 1$. In our considerations, we assume that the end of the inflation is indicated by the condition $\epsilon_H = 1$.

At same time when the velocity of the inflaton grows large enough to go beyond the slow-roll regime, a~negative value of field space curvature can cause so-called geometrical destabilization \cite{Renaux-Petel:2015mga} of perturbations of the field $\chi$. Therefore, in order to track the dynamics of perturbations accurately, both fields should be taken into account around the end of the inflation. In Section \ref{sec:first_order_perturbation} we will briefly present linear analysis of perturbations of inflaton $\phi$ and spectator field $\chi$. Our results of nonlinear numerical lattice simulations will be described in Section \ref{sec:results_of_numerical_simulations}.

\subsection{First-order perturbations\label{sec:first_order_perturbation}}

Equations of motion for perturbations in two-field models described by (\ref{eq:field_space_metric}) can be found e.g.\ in \cite{Lalak:2007vi}, without slow-roll approximation or any additional assumptions. The perturbed Friedmann-Robertson-Walker metric (in longitudinal gauge, with only scalar degrees of freedom included and constraints taken into account) reads
\begin{equation}
\mathrm{d}s^2=-(1+2\Psi)\,\mathrm{d}t^2+a^2(1-2\Psi)\,{\mathrm{d}\mathbf{x}^2}.
\end{equation}
Linear perturbations are described in terms of gauge-invariant Mukhanov-Sasaki variables:
\begin{align}
Q_\phi &:= \delta\phi+\frac{\dot{\phi}}{H}\Psi && \textrm{and} & Q_\chi &:= \delta\chi+\frac{\dot{\chi}}{H}\Psi,
\end{align}
which, for $\chi=0$, obey very similar equations of motion
\begin{equation}
\ddot{Q}_{\phi / \chi} + 3H\dot{Q}_{\phi/\chi} + \left(\frac{k^2}{a^2} + m^2_{\phi/\chi}\right)Q_{\phi/\chi} = 0, \label{eq:eom_perturbations}
\end{equation}
with the only difference in effective masses
\begin{align}
m^2_\phi &= V_{\phi\phi}, & m_\chi^2 &= V_{\chi\chi}+\frac{1}{2}\dot{\phi}^2\mathbb{R}.\label{eq:m2_perturbations}
\end{align}
Writing eqs. \eqref{eq:eom_perturbations} and \eqref{eq:m2_perturbations}, we neglected all contributions suppressed by the Planck scale, as the energy scale of inflation is much smaller.
We also used the assumption $\chi=0$ which immediately implies that $V_{\chi}(\phi,0)=0$ and  $V_{\phi\chi}(\phi,0)=0$ for $V$ given by \eqref{eq:potential}.

Taking into account the fact that the field space curvature is negative, according to eq.~\eqref{eq:field_space_curvature}, one deduces from eq.~\eqref{eq:m2_perturbations} that for large values of $\beta$, i.e.\ for small values of $\alpha$, perturbations $Q_\chi$ are expected to exhibit an \emph{intermittent tachyonic instability} as $|\dot{\phi}|$ increases towards the end of inflation. The linear analysis of evolution of perturbations of both the inflaton $Q_\phi$ and the spectator $Q_\chi$ during preheating, performed on the basis of the Floquet theorem in \cite{Krajewski:2018moi} and extended in \cite{Iarygina:2018kee}, confirmed this prediction. Moreover, the so-called Floquet exponents, which measure speed of the exponential growth of perturbations during instability in oscillating background of the inflaton $\phi$, turned out to be much larger for the spectator than for the inflaton, revealing that the described \emph{intermittent tachyonic instability} is an~important factor for mode amplification and particle production, thereby competing with the parametric resonance of the inflaton perturbations, which was found to be effective for reheating even for single-field model \cite{Lozanov:2016hid, Lozanov:2017hjm}. 

In \cite{Iarygina:2018kee}, the authors described an intriguing simple scaling of the instability of linear spectator perturbations $Q_\chi$ (especially the Floquet exponents) with changes of $\alpha$ in regime of high curvature of the field space. Their numerical calculation of the Floquet exponents as functions of $\phi_0 / \sqrt{\alpha}$ and $k / \mu = k \sqrt{\alpha} M_{Pl} / M^2$
give very similar results 
for $\alpha = 10^{-3}, 10^{-4}$;
those results are also quite similar for those for $\alpha=10^{-2}$.

Analytic arguments explaining the observed universality can be briefly
summarized, as follows. 
The potential \eqref{eq:potential} is a~fixed function of $\delta = \phi / \sqrt{\alpha}$ for $\chi=0$. The calculation in slow-roll approximation shows that the fraction $\delta_\text{end} = \phi_\text{end} / \sqrt{\alpha}$ of the value of the inflaton field at the end of inflation $\phi_\text{end}$ (and the beginning of oscillations) has only logarithmic dependence on $\alpha$ (and on $n$). In fact, this dependence leads to residual one of other quantities. Furthermore, it turns out that the value of parameter $\mu = M^2 / \sqrt{\alpha}$ is determined independently of $\alpha$ from normalization of scalar power spectrum. Hence, the dynamics of background oscillations of $\delta$ is independent of $\alpha$ up to dependence of Hubble parameter $H$ which can be neglected in the first approximation since frequency of oscillations $\omega$ for small $\alpha$ is much higher then Hubble time. As a result, the oscillations of $\delta$ have the amplitude which is logarithmically dependent on $\alpha$ and they are weakly damped (proportionally to $H$). This leads to conclusion that the effective mass $m_\chi^2$ of the spectator perturbations is 
practically independent of~$\alpha$ when expressed in terms of $\delta$ and $\mu$, since the dependence on $\alpha$ cancels out between $\mathbb{R}$ and $\dot{\phi}^2$ (similarly, dependence from derivatives with respect to $\chi$ cancels the one from $M^4$ in $V_{\chi \chi}$ term). Neglecting the time dependence of the scale factor $a$ one obtains the Floquet exponents expressed as functions of $\delta_0$ and $\mu$, without an explicit
dependence on~$\alpha$.

The dependence of the Floquet exponents on potential parameter $n$ is also discussed in \cite{Iarygina:2018kee}, but it is not as straightforward as the dependence on $\alpha$. The similarities are visible only for highly curved field spaces with $\alpha \lesssim 10^{-4}$ when the reheating is essentially instantaneous (and with $\phi / \phi_\text{end}$ used instead of $\delta$).

The authors of \cite{Iarygina:2018kee} also estimated the duration of  reheating, calculating the time required to transfer the entire inflaton energy density into the energy of spectator perturbations, extrapolating linear analysis outside its validity range. 
Therefore, that estimation has obvious limitations.
The perturbative expansion will break down as soon as the produced fluctuations will backreact on the background evolution. 
Moreover, the modes with large wavenumbers will be produced in re-scatterings due to nonlinearity of the potential \eqref{eq:potential} leading to broadening of the energy spectrum of fluctuations. These effects can be studied to some extent in numerical lattice simulations \cite{Felder:2000hj, Felder:2001kt}, as we discuss in Section \ref{sec:results_of_numerical_simulations}.
 
\section{Numerical lattice simulations}
\label{sec:three}

\subsection{Description of methods}

We developed own code directly designed for numerical studies 
of the evolution 
of scalar perturbations in $\alpha$-attractor models
on the lattice.
Our main rationale was that the existing codes written to simulate preheating after inflation on the lattice  (see e.g~\cite{Amin:2014eta} for a~review) are not suitable for non-canonical kinetic terms. The only exception is GABE, which is, however, non-symplecic
and thus poorly reproduce energy-conservation.

The theoretical background of our approach is described in detail in the appendix A of \cite{Krajewski:2018moi}. We used second order central finite difference scheme to discretize in space the continuum theory given by the Lagrangian density \eqref{eq:field_space_metric}. Introducing canonical momenta $\pi_\phi$, $\pi_\chi$, $p_a$ conjugate to $\phi$, $\chi$ and $a$, respectively, we formed a discrete Hamiltonian describing the dynamics of finite number of degrees of freedom in the discretized theory. It turns out that such a discrete Hamiltonian (analogously as its continuum counterpart) is separable and can be divided into four terms in such a~way that none of the parts depends on both the canonical variable and its conjugate momentum. Hamilton's equation for each term separately can be integrated exactly. Using collocation method for these four exact flows we formed symplectic integrator of a second-order precision in time. 

In order to control the error made during numerical computation and verify the reliability of results we used an~error indicator based on a~priori error estimation to detect the breakdown of used time integration scheme. Similar technique was successfully used by one of us in the past \cite{Krajewski:2016vbr, Krajewski:2017czs, Krajewski:2019vix, Krajewski:2021jje} to control time step length in simulations of dynamics of cosmological domain walls. The rapid growth of the numerical estimate of the leading term in the expansion of the error of time integration in powers of time step displays failure of the used numerical scheme in resolving dynamics of the physical problem. It can be connected to production of fluctuations with wavelengths shorter then lattice spacing $h$ (i.e. physical distant between neighbouring lattice points) which can not be represented properly on constant lattice and whose dynamics take place at shorter time scales. Moreover, we performed 5 independent simulations for each set of parameters which we compared and verified their consistency. This attempt guarantee full reliability of the presented results in contrast to consistency verification alone which cannot detect some systematic error.\footnote{In addition we unit tested our numerical code, thus the existence of unknown bugs in it is unlikely.} 

The main development in our code with respect to \cite{Krajewski:2018moi} is the usage of GPU acceleration to speed up our computations, allowing us to increase the size of lattice from $128^3$ used previously to $512^3$. Taking advantage of the obtained decrease in computation time, 
we were able to study deeper the parametric dependence of geometrical destabilisation. 
We simulated the process of the preheating for 12 different parameters sets, with $\alpha=10^{-3}$, $10^{-3.5}$ and $10^{-4}$, and $n=1, 1.5, 2$ and $3$. 
For each set of $(\alpha,n),$ we performed simulations with three different lattice spacings, which we parametrize by the maximal wavevector $k_\text{max} = \sqrt{3} \pi / h$ that fits on a cubic lattice of spacing~$h$. 
Comparing results of simulations with values of $k_\text{max}$ that differ by the factor of $2$ we were able to study the effects of spatial discretization of equations of motion and the role of IR and UV cutoffs that are inevitably associated with lattice simulations. 
For each parameter set, we chose a value of~$k_\text{max}$ that ensured a~good trade-off between granularity in the Floquet instability regions and factoring in higher frequency modes. Then, simulations with lower cutoff confirm that the production of fluctuations due to first instability band is properly reproduced and ones with higher cutoff that modes with wavelengths lower than UV cutoff do not have significant influence on the evolution during simulated period of time.

The evolution of fields $\phi$ and $\chi$ obtained from our lattice
simulations was also used
to determine the spectrum of gravitational waves emitted during (p)reheating. The details of this calculation are described in Section \ref{sec:GWs_spectrum}.

\subsection{Initial conditions for perturbations}
At this point we take an~advantage of the eq. \eqref{eq:eom_perturbations} and describe initial conditions for our numerical simulations.

One usually chooses the Bunch-Davies initial conditions for perturbations which are appropriate for quantum fields in time-dependent, de Sitter background. This procedure is well-known for fields with trivial field-space metric and was generalized for non-trivial cases, e.g. for two-field models with kinetic terms as in the eq.~\eqref{eq:field_space_metric}, the prescription for initial conditions of the perturbations can be found e.g.\ in \cite{Lalak:2007vi}. It simplifies when $\chi=0$ and when one can neglect the contributions suppressed by the Planck scale.

Defining $u_\phi := aQ_\phi$ and $u_\chi := aQ_\chi$, and using conformal time $\tau$ (related to the cosmic time $t$ as $\mathrm{d}\tau = a^{-1} \mathrm{d}t$), eqs.~\eqref{eq:eom_perturbations} can be written as equations of motion of two independent harmonic oscillators with time dependent masses (for details, see e.g.~\cite{Langlois:2008mn}). Therefore, a usual single-field quantization procedure (see, e.g.~\cite{Mukhanov:2007}) can be performed. In the so-called adiabatic approximation, it provides the following initial conditions for perturbations at $\tau = \tau_0$:
\begin{align}
u_I(k,\tau_0) &= \frac{1}{\sqrt{2\omega_{I,k}}}\textrm{e}^{-i\omega_{I,k}\tau_0}, & u'_{I}(k,\tau_0) =&-i\sqrt{\frac{\omega_{I,k}}{2}}\textrm{e}^{-i\omega_{I,k}\tau_0}, \label{eq:initial_conditions_for_perturabtions}
\end{align} 
where $I\in\{\phi,\chi\}$, the prime ${}'$ denotes a derivative with respect to the conformal time and the frequencies $\omega_{I,k}$ are defines as:  
\begin{equation}
\omega_{I,k}^2 = k^2+a^2\left(m^2_{I} + a^{-3} \partial_\tau^2 a \right) \approx k^2+a^2\left(m^2_{I} + 2H^2\right),
\end{equation}
where in the last approximate equality, the terms proportional to slow-roll parameters were neglected.

We use expressions \eqref{eq:initial_conditions_for_perturabtions} to set Gaussian initial conditions for perturbations in our lattice simulations in the similar way as it is done for LATTICEEASY code \cite{Felder:2000hq} (however later we transform velocities into conjugate momenta).

The adiabatic approximation can be used when the following condition
is satisfied:
\begin{equation}
\frac{\omega_I^{-1}}{2}\frac{\mathrm{d}^2\omega_I^{-1}}{\mathrm{d}\tau^2} - 
\frac{1}{4}\left( \frac{\mathrm{d}\omega_I^{-1}}{\mathrm{d}\tau}\right)^2 \ll 1 \, .
\label{eq:adiabacity_contidion}
\end{equation}
The initial time $\tau_0$ for the lattice simulations must be chosen so that the condition \eqref{eq:adiabacity_contidion} holds true.

\subsection{Results of simulations\label{sec:results_of_numerical_simulations}}
We start with the description of results of simulations with $n=1$ and $n=1.5$, and for $\alpha$ ranging from $10^{-4}$ to $10^{-3}$, since for those sets of parameters, lattice simulations were previously performed by us and discussed in \cite{Krajewski:2018moi}.

From the point of view of comparing CMB measurements with predictions of 
inflationary models, the barotropic parameter $w=p/\rho$ is a very useful quantity, as it is directly connected to the way the Universe expands. 
In our model, the barotorpic parameter can be written in terms of fields as:
\begin{equation}
w = \frac{\frac{1}{2}\left(\textrm{e}^{2b(\chi)}\dot{\phi}^2+\dot{\chi}^2\right) -\frac{1}{6a^2}\left(\textrm{e}^{2b(\chi)}(\nabla\phi)^2+(\nabla\chi)^2\right)-V(\phi,\chi)}{\frac{1}{2}\left(\textrm{e}^{2b(\chi)}\dot{\phi}^2+\dot{\chi}^2\right)+\frac{1}{2a^2}\left(\textrm{e}^{2b(\chi)}(\nabla\phi)^2+(\nabla\chi)^2\right)+V(\phi,\chi)}\,, \label{eq:definition_barotropic_parameter}
\end{equation}
where spatial averaging is assumed for both the numerator 
and the denominator.

The evolution of $w$ is presented in figure \ref{fig:n=1,1.5_plot_barotropic} for $n=1$ and $n=1.5$. It can be seen that the value of barotropic parameter oscillates, which begins within the end of the inflation, corresponding to $N=0$, due to the periodic transfer of the energy between the kinetic and the potential energy of the inflaton. The initial amplitude of the oscillations is from $-1$ to $1$. However, due to instability the amplitude decreases, because the energy of the homogeneous background is transferred into fluctuations with gradient component of the energy. For all investigated sets of parameters, 
this decrease is visible and, with the exception of $n=1$ and $\alpha=10^{-3}$, it starts rapidly after a~few oscillations 
of the background inflaton field, in just a~fraction of the e-fold after the end of the inflation. As we argued in \cite{Krajewski:2018moi}, and confirm in the present paper, rapid evolution of the barotropic parameter 
is caused by efficient production of fluctuations of fields due to geometrical destabilization.

\begin{figure*}[!ht]
	\begin{subfigure}[t]{0.5\textwidth}
		\label{fig:parameters_alpha=1e-3_n=1_plot_barotropic}
	    \flushleft
		\includegraphics[width=215pt]{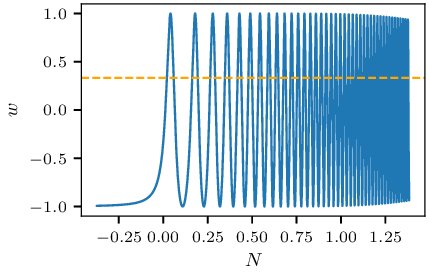}
	\end{subfigure} \hfill %
	\begin{subfigure}[t]{0.5\textwidth}
		\label{fig:parameters_alpha=1e-3_n=1.5_plot_barotropic}
		\flushright
		\includegraphics[width=215pt]{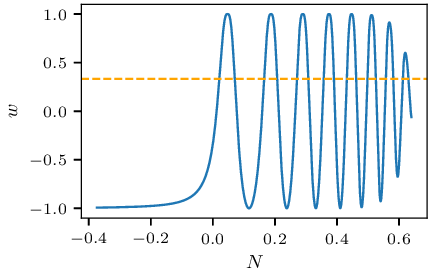}
	\end{subfigure} \\
	\begin{subfigure}[t]{0.5\textwidth}
		\label{fig:parameters_alpha=1e-3.5_n=1_plot_barotropic}
		\flushleft
		\includegraphics[width=215pt]{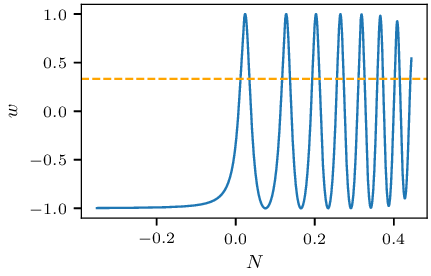}
	\end{subfigure}\hfill %
	\begin{subfigure}[t]{0.5\textwidth}
		\label{fig:parameters_alpha=1e-3.5_n=1.5_plot_barotropic}
		\flushright
		\includegraphics[width=215pt]{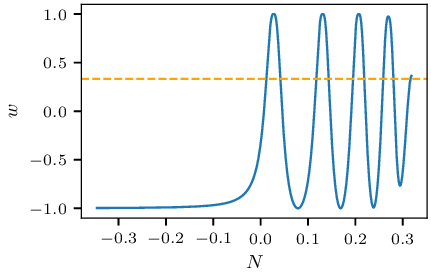}
	\end{subfigure}\\
	\begin{subfigure}[t]{0.5\textwidth}
		\label{fig:parameters_alpha=1e-4_n=1_plot_barotropic}
	    \flushleft
		\includegraphics[width=215pt]{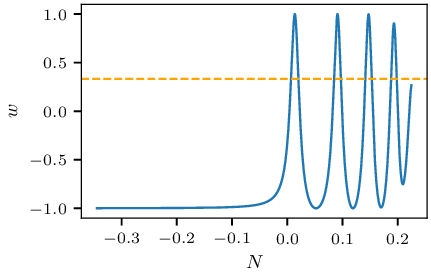}
	\end{subfigure}\hfill %
	\begin{subfigure}[t]{0.5\textwidth}
		\label{fig:parameters_alpha=1e-4_n=1.5_plot_barotropic}
		\flushright
		\includegraphics[width=215pt]{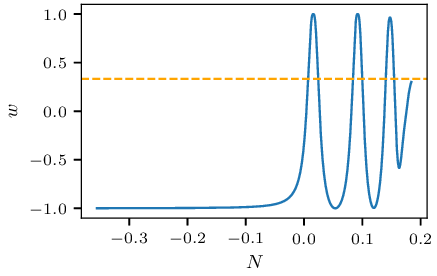}
	\end{subfigure} %
	\caption{Evolution of the barotropic parameter $w$ (blue solid line) as a function of the number of e-folds $N$ from the end of inflation ($\epsilon = 1$) for $n=1$ (left column) and $n=1.5$ (right column) and $\alpha = 10^{-3}$ (top), $\alpha=10^{-3.5}$ (middle), $\alpha=10^{-4}$ (bottom). Dashed orange vertical line correspond to $w=1/3$. \protect\label{fig:n=1,1.5_plot_barotropic}}
\end{figure*}

Comparing figure~\ref{fig:n=1,1.5_plot_barotropic} with
the figure 6 in \cite{Krajewski:2018moi}, it can be seen that 
the improved spatial resolution allows for longer and
more accurate simulations, but our lattice size is not yet
sufficient to proclaim the fate of the barotropic parameter
after inflation, as we are unable to determine the asymptotic
value of~$w$. 

More insight about this matter can be gained
from studying the evolution of various components of the energy density,
which we present in figure~\ref{fig:n=1,1.5_plot_energy}. 
The initial evolution of energy density is the same as in single-field models. From the beginning of the simulation, the potential energy $E_{pot}$ decreases slowly, translating into the kinetic energy $E_{kin, \phi}$ of the inflaton. From the end of inflation, the potential energy density starts to oscillate and the energy is exchanged between the potential and the kinetic components. However, around that time, geometrical destabilization becomes important and both the kinetic $E_{kin, \chi}$ and gradient $E_{grad, \chi}$ energy densities of the spectator start increasing. Such a simultaneous growth at similar rate 
indicates that mainly the spatial fluctuations of the spectator are produced and the production of the homogeneous mode is subdominant. Finally, after a~fraction of an e-fold after the end of inflation, the gradient energy density $E_{grad, \phi}$ of the inflaton starts to grow rapidly. This is a~symptom of the production of fluctuations of the inflaton caused by the nonlinear interactions with the induced modes of the spectator field, the effect that cannot be deduced from linear Floquet analysis. During further evolution, the energy density stored in fluctuations of both fields increases. 
With the exception of $n=1$ with $\alpha=10^{-3}$, we can follow the
evolution of the fields until the kinetic and gradient energy density
of the spectator is within one order of magnitude from the
average energy density of the homogeneous inflaton field. 
Comparing the growth of the energy in spectator fluctuations to the growth
of the energy in inflaton fluctuations in the absence of the spectator,
determined in \cite{Lozanov:2016hid}, we conclude that in the full
two-field model the instability is much stronger, as it takes a fraction of an e-fold to transfer the energy density from the homogeneous mode to the fluctuations, instead of a few e-folds \cite{Lozanov:2016hid}.

Unfortunately, the investigated instability is so strong that the used discretization of equations of motion on the constant lattice breaks down due to production of short wavelength modes (shorter than the lattice spacing $h$) and we have to stop our simulations around the time when the energy density of fluctuations is dominant component since the further calculations would give unreliable results. 

We approach this breakdown of the simulations in a two-step way.
First, we keep track of the amplitude of the model with $k=k_{\rm max}$.
When this amplitude grows 5 times with respect to its initial value,
we treat the results obtained for later times as potentially unreliable,
because the power in the fluctuations should UV cascade to smaller scales,
which we are unable to resolve. 
The amplitude of the fluctuations can be overestimated, 
since the modes with wavenumbers larger than the UV cutoff 
cannot be simulated on the lattice; 
the energy that should be deposited in those modes through rescattering 
stays with lower-$k$ modes.

Therefore, in our presentation of numerical results for the amplitudes of the fluctuations we clearly mark
later times by denoting respective curves with grey color. However, 
we have to note that adopting this conservative approach does not mean
that the amplitudes of fluctuations cease to grow, we only lack certainty
about the rate of this growth.

In addition to physical breakdown of our simulations, 
resulting from the fact that
the lattice approach allows us to simulate only a finite volume of
space with a finite resolution, there is also a numerical breakdown,
when the error becomes unacceptably large. In our calculations, 
the latter breakdown always follows the former.
Yet even with the conservative approach adopted herein,
our calculations clearly demonstrate that 
very shortly after the simulated phase of the evolution
of the Universe the single-field description of the considered model
is not valid due to the non-trivial dynamics of the spectator.

\begin{figure*}[!ht]
	\begin{subfigure}[t]{0.5\textwidth}
		\label{fig:parameters_alpha=1e-3_n=1_plot_energy}
	    \flushleft
		\includegraphics[width=215pt]{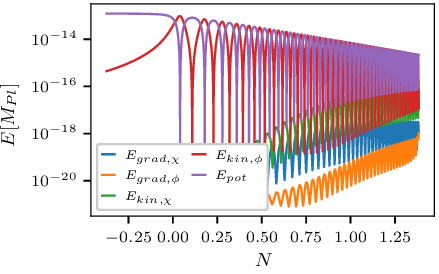}
	\end{subfigure} \hfill %
	\begin{subfigure}[t]{0.5\textwidth}
		\label{fig:parameters_alpha=1e-3_n=1.5_plot_energy}
		\flushright
		\includegraphics[width=215pt]{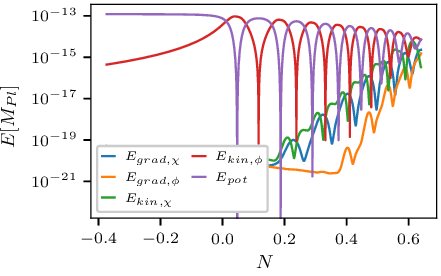}
	\end{subfigure} \\
	\begin{subfigure}[t]{0.5\textwidth}
		\label{fig:parameters_alpha=1e-3.5_n=1_plot_energy}
		\flushleft
		\includegraphics[width=215pt]{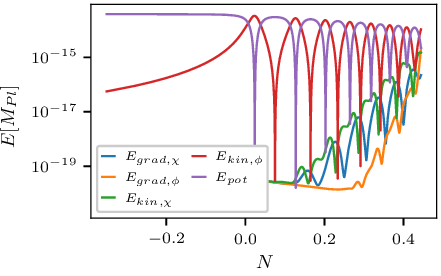}
	\end{subfigure}\hfill %
	\begin{subfigure}[t]{0.5\textwidth}
		\label{fig:parameters_alpha=1e-3.5_n=1.5_plot_energy}
		\flushright
		\includegraphics[width=215pt]{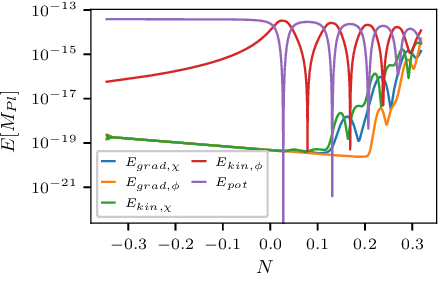}
	\end{subfigure}\\
	\begin{subfigure}[t]{0.5\textwidth}
		\label{fig:parameters_alpha=1e-4_n=1_plot_energy}
	    \flushleft
		\includegraphics[width=215pt]{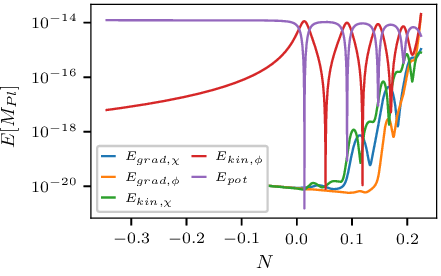}
	\end{subfigure}\hfill %
	\begin{subfigure}[t]{0.5\textwidth}
		\label{fig:parameters_alpha=1e-4_n=1.5_plot_energy}
		\flushright
		\includegraphics[width=215pt]{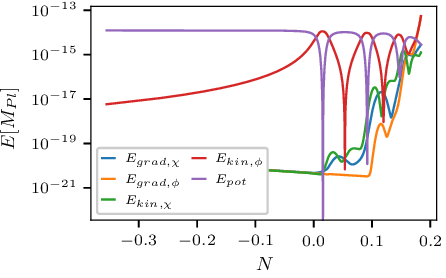}
	\end{subfigure} %
	\caption{Evolution of energy components as a function of the number of e-folds $N$ from the end of inflation for $n=1$ (left column) and $n=1.5$ (right column) and $\alpha = 10^{-3}$ (top), $\alpha=10^{-3.5}$ (middle), $\alpha=10^{-4}$ (bottom). \protect\label{fig:n=1,1.5_plot_energy}}
\end{figure*}

The evolution of fluctuations of fields can be also studied with Fourier transform of fields strength. In figure~\ref{fig:n=1,1.5_plot_chi_fluct_k}, 
we present the evolution of amplitudes of the Fourier transform of the spectator field $\chi$ for selected wave vectors $k$. Simultaneously with the end of inflation the amplitude of the lowest $k$ mode available on the lattice (deep red) starts to grow by orders of magnitude. On the other hand, high $k$ modes stay nearly unaffected for some time until they rapidly increase close to the end of simulation. Modes for some intermediate values of $k$ develop in between.

\begin{figure*}[!ht]
	\begin{subfigure}[t]{0.5\textwidth}
		\label{fig:parameters_alpha=1e-3_n=1_plot_chi_fluct_k}
	    \flushleft
		\includegraphics[width=215pt]{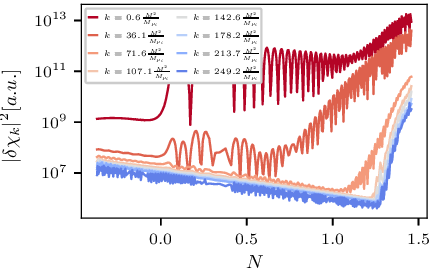}
	\end{subfigure} \hfill %
	\begin{subfigure}[t]{0.5\textwidth}
		\label{fig:parameters_alpha=1e-3_n=1.5_plot_chi_fluct_k}
		\flushright
		\includegraphics[width=215pt]{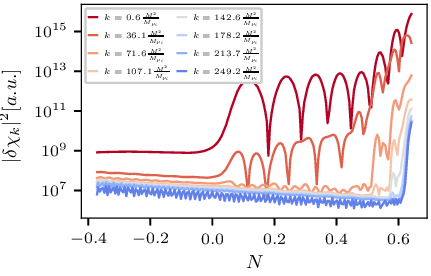}
	\end{subfigure} \\
	\begin{subfigure}[t]{0.5\textwidth}
		\label{fig:parameters_alpha=1e-3.5_n=1_plot_chi_fluct_k}
		\flushleft
		\includegraphics[width=215pt]{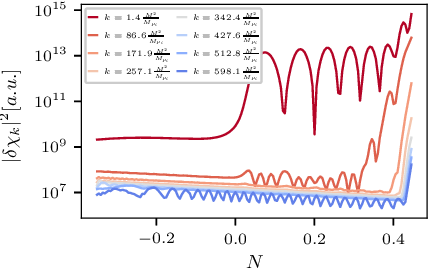}
	\end{subfigure}\hfill %
	\begin{subfigure}[t]{0.5\textwidth}
		\label{fig:parameters_alpha=1e-3.5_n=1.5_plot_chi_fluct_k}
		\flushright
		\includegraphics[width=215pt]{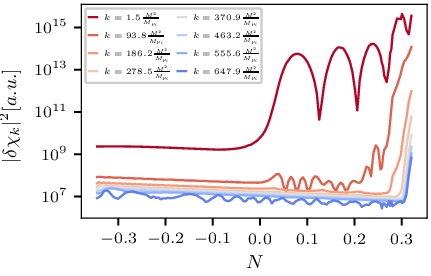}
	\end{subfigure}\\
	\begin{subfigure}[t]{0.5\textwidth}
		\label{fig:parameters_alpha=1e-4_n=1_plot_chi_fluct_k}
	    \flushleft
		\includegraphics[width=215pt]{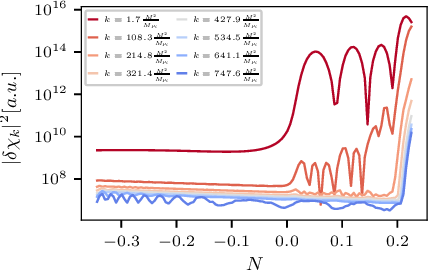}
	\end{subfigure}\hfill %
	\begin{subfigure}[t]{0.5\textwidth}
		\label{fig:parameters_alpha=1e-4_n=1.5_plot_chi_fluct_k}
		\flushright
		\includegraphics[width=215pt]{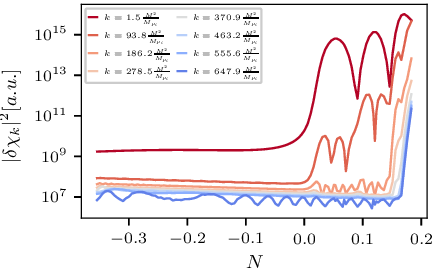}
	\end{subfigure} %
	\caption{Evolution of amplitudes of selected modes of fluctuations $\delta \chi$ of the spectator field as a function of the number of e-folds $N$ from the end of inflation ($\epsilon = 1$) for $n=1$ (left column) and $n=1.5$ (right column) and $\alpha = 10^{-3}$ (top), $\alpha=10^{-3.5}$ (middle), $\alpha=10^{-4}$ (bottom). \protect\label{fig:n=1,1.5_plot_chi_fluct_k}}
\end{figure*}

Some further characteristics of the instability can be recognized from  spectra of fluctuations at various fixed times. We present those spectra in figure \ref{fig:n=1,1.5_plot_chi_fluct_N}. We can see that at the beginning of the destabilization only modes with wave vectors $k$ smaller than the scale of the negative mass squared of the spectator are excited. Such pattern is predicted by the linear Floquet analysis, because these modes are excited by the tachyonic instability caused by the negative contribution to the effective mass coming from negative curvature of the field space. Moreover, a `ringing' pattern resulting from rescattering of amplified modes is clearly visible. Near the end of our simulations, these low-$k$ modes produced due to tachyonic instability backreact and cause the growth of modes with larger wave vector $k$ including finally all modes available on the lattice. In contrast to the initial stage of the destabilization, which  can be modelled via linear approximation, amplification of high $k$ modes is a~purely nonlinear effect. However, this effect plays a~very important role in reheating, since it leads to the fragmentation of fields and,
consequently, to the transition to the radiation domination era. It is worth stressing that the maxima of the spectra of spectator fluctuations for all simulations are well separated from the $k=0$, i.e. the value at the maximum is always orders of magnitude higher than the value for the lowest $k$ available on the lattice. This proves that the dynamics of low-$k$ modes is reliably reproduced in all our simulations during the considered period of time, thus the IR cutoff is well chosen.

\begin{figure*}[!ht]
	\begin{subfigure}[t]{0.5\textwidth}
		\label{fig:parameters_alpha=1e-3_n=1_plot_chi_fluct_N}
	    \flushleft
		\includegraphics[width=215pt]{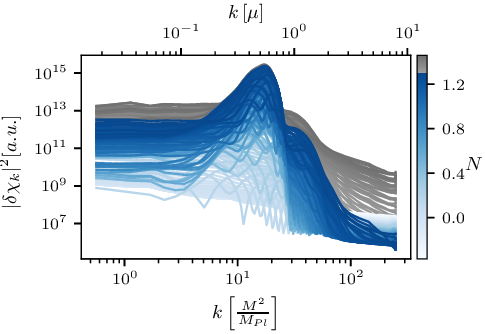}
	\end{subfigure} \hfill %
	\begin{subfigure}[t]{0.5\textwidth}
		\label{fig:parameters_alpha=1e-3_n=1.5_plot_chi_fluct_N}
		\flushright
		\includegraphics[width=215pt]{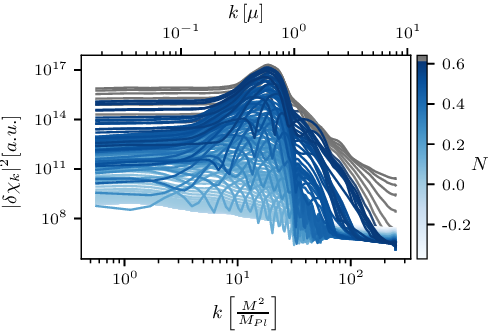}
	\end{subfigure} \\
	\begin{subfigure}[t]{0.5\textwidth}
		\label{fig:parameters_alpha=1e-3.5_n=1_plot_chi_fluct_N}
		\flushleft
		\includegraphics[width=215pt]{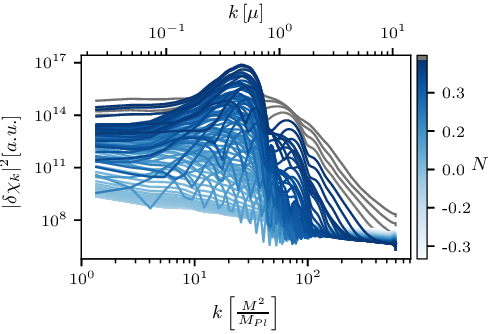}
	\end{subfigure}\hfill %
	\begin{subfigure}[t]{0.5\textwidth}
		\label{fig:parameters_alpha=1e-3.5_n=1.5_plot_chi_fluct_N}
		\flushright
		\includegraphics[width=215pt]{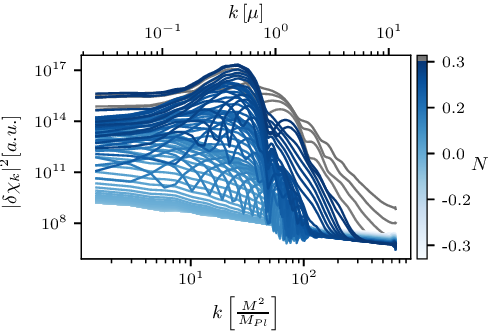}
	\end{subfigure}\\
	\begin{subfigure}[t]{0.5\textwidth}
		\label{fig:parameters_alpha=1e-4_n=1_plot_chi_fluct_N}
	    \flushleft
		\includegraphics[width=215pt]{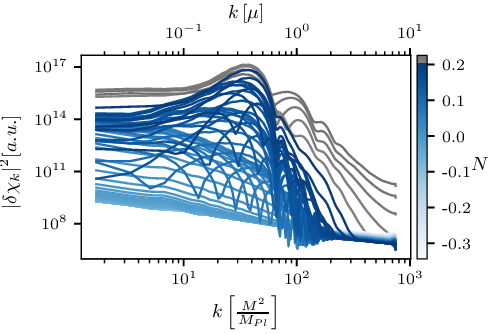}
	\end{subfigure}\hfill %
	\begin{subfigure}[t]{0.5\textwidth}
		\label{fig:parameters_alpha=1e-4_n=1.5_plot_chi_fluct_N}
		\flushright
		\includegraphics[width=215pt]{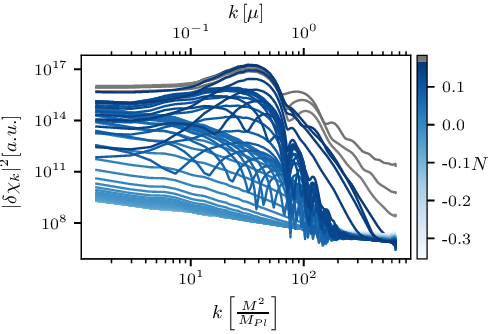}
	\end{subfigure} %
	\caption{Evolution of the spectrum of fluctuations $\delta \chi$ of the spectator field as a function of the number of e-folds $N$ from the end of inflation for $n=1$ (left column) and $n=1.5$ (right column) and $\alpha = 10^{-3}$ (top), $\alpha=10^{-3.5}$ (middle), $\alpha=10^{-4}$ (bottom).
	Lines of different colors correspond to different time points
	described by the number of e-folds after the end of inflation,
	as indicated on the bar next to each panel. Grey lines denote
    results obtained after a potential breakdown of the simulations, as discussed in the text. Lines are drawn for time steps corresponding to 0.005 efolds.
	\protect\label{fig:n=1,1.5_plot_chi_fluct_N}}
\end{figure*}

For completeness, we also show analogous plots for fluctuations $\delta \phi$ of the inflaton in figure \ref{fig:n=1,1.5_plot_phi_fluct_N}. 
Again, we can see for most of the simulated period of time, only low-$k$ modes of the inflaton fluctuations are excited. This observation is consistent with the Floquet analysis, since the inflaton is affected only by the parametric resonance \cite{Lozanov:2016hid, Lozanov:2017hjm, Krajewski:2018moi}, 
which is much weaker than the tachyonic instability of the spectator. However, towards the end of the simulation higher wave vector modes of $\delta \phi$ follow those of $\delta \chi$, since the two fields $\phi$ and $\chi$ are tightly coupled through the non-canonical kinetic term. Although the amplitudes of modes of the fluctuations $\delta \phi$ of the inflaton are smaller, the contributions to the energy density coming from them dominate over those of $ \delta \chi$ due to the factor of $\cosh^2(\beta\chi)$ multiplying the kinetic term of $\phi$. 

\begin{figure*}[!ht]
	\begin{subfigure}[t]{0.5\textwidth}
		\label{fig:parameters_alpha=1e-3_n=1_plot_phi_fluct_N}
	    \flushleft
		\includegraphics[width=215pt]{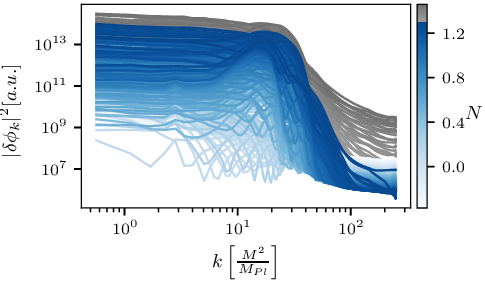}
	\end{subfigure} \hfill %
	\begin{subfigure}[t]{0.5\textwidth}
		\label{fig:parameters_alpha=1e-3_n=1.5_plot_phi_fluct_N}
		\flushright
		\includegraphics[width=215pt]{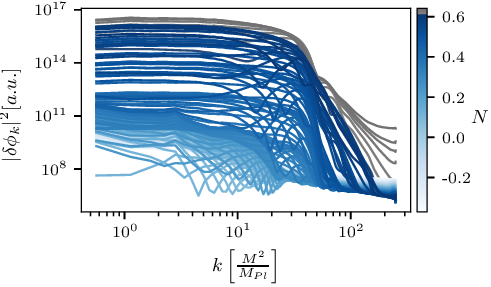}
	\end{subfigure} \\
	\begin{subfigure}[t]{0.5\textwidth}
		\label{fig:parameters_alpha=1e-3.5_n=1_plot_phi_fluct_N}
		\flushleft
		\includegraphics[width=215pt]{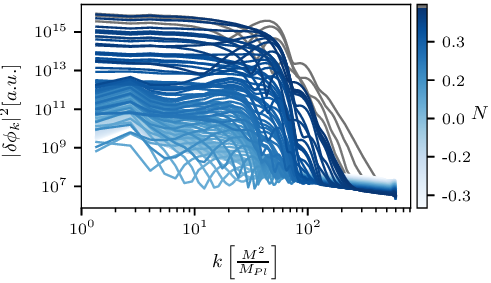}
	\end{subfigure}\hfill %
	\begin{subfigure}[t]{0.5\textwidth}
		\label{fig:parameters_alpha=1e-3.5_n=1.5_plot_phi_fluct_N}
		\flushright
		\includegraphics[width=215pt]{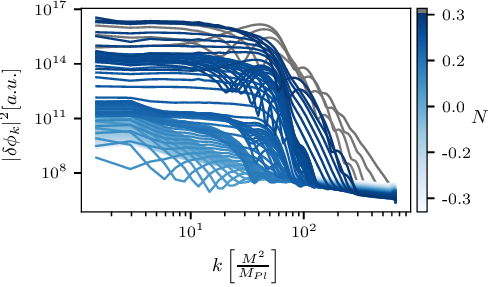}
	\end{subfigure}\\
	\begin{subfigure}[t]{0.5\textwidth}
		\label{fig:parameters_alpha=1e-4_n=1_plot_phi_fluct_N}
	    \flushleft
		\includegraphics[width=215pt]{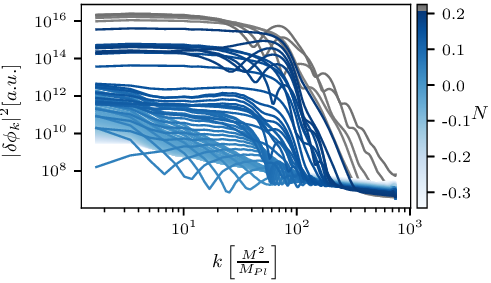}
	\end{subfigure}\hfill %
	\begin{subfigure}[t]{0.5\textwidth}
		\label{fig:parameters_alpha=1e-4_n=1.5_plot_phi_fluct_N}
		\flushright
		\includegraphics[width=215pt]{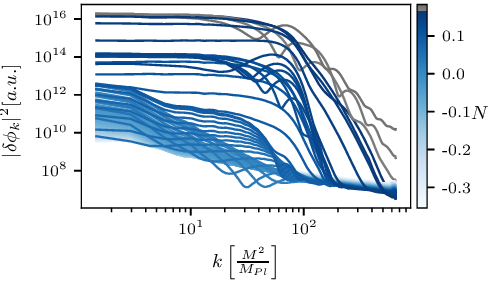}
	\end{subfigure} %
	\caption{Evolution of the spectrum of fluctuations $\delta \phi$ of the spectator field as a function of the number of e-folds $N$ from the end of inflation for $n=1$ (left column) and $n=1.5$ (right column) and $\alpha = 10^{-3}$ (top), $\alpha=10^{-3.5}$ (middle), $\alpha=10^{-4}$ (bottom).
	Lines of different colors correspond to different time points
	described by the number of e-folds after the end of inflation,
	as indicated on the bar next to each panel. Grey lines denote
    results obtained after a potential breakdown of the simulations, as discussed in the text. Lines are drawn for time steps corresponding to 0.005 efolds. \protect\label{fig:n=1,1.5_plot_phi_fluct_N}}
\end{figure*}

So far, we have discussed the dynamics of geometrical destabilization at the end of inflation, 
taking as examples models with $n=1$ and $n=1.5$.
Now we shall also address models with $n=2$ and $n=3$, in part to study
the parameter dependence of that phenomenon.
In figure~\ref{fig:n=2,3_plot_barotropic}, we show the evolution of the barotropic parameter for models with $n=2$ and $n=3$. 
Comparing these plots with those in figure~\ref{fig:n=1,1.5_plot_barotropic},
we see that decreasing the value of $\alpha$ parameter decreases the number of e-folds needed for a prospective transition to radiation-domination era. 
With the notable exception of the case $n=1$, $\alpha=10^{-3}$, 
this number of e-folds is smaller than $1$ for $\alpha \le 10^{-3}$; thus we are corroborating our claim from \cite{Krajewski:2018moi} 
that for small values of $\alpha\le 10^{-3}$, reheating in two-field T-models is practically instantaneous. 
The duration of reheating decreases with increasing value of $n$. In all cases, a shorter transition time is associated with a decreasing number of oscillations of background inflaton field needed to transfer energy to fluctuations.\footnote{The number of oscillations of the inflaton can be directly counted from zeros of potential energy density in figures \ref{fig:n=1,1.5_plot_energy} and \ref{fig:n=2,3_plot_energy}. However, each oscillation before the transition is associated with the oscillation of barotropic parameter $w$.} These observations are in agreement with predictions of approximate analysis of ref. \cite{Iarygina:2018kee}. 
This apparent concordance hints at a certain subtlety: in our
simulations we can see the onset of nonlinear backreaction,
which, in particular, leads to excitations of the inflaton
modes through rescattering and is vital for the evolution of the
barotropic parameter. The oscillations of~$w$ have a reduced amplitude
towards the end of our simulation, which indicates that~$w$ settles
at some values. Unfortunately, the resolution of our
simulations is insufficient to determine whether
the asymptotic value is $w=1/3$, characteristic of radiation.
We would like to stress that such tendency could not
be reliably foreseen from approximate, linear analyses.

\begin{figure*}[!ht]
	\begin{subfigure}[t]{0.5\textwidth}
		\label{fig:parameters_alpha=1e-3_n=2_plot_barotropic}
		\flushleft
		\includegraphics[width=215pt]{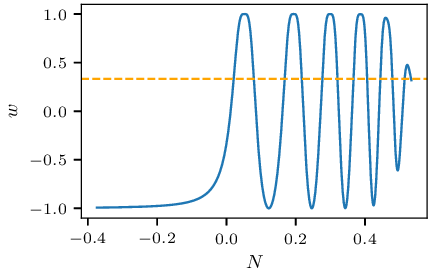}
	\end{subfigure}\hfill %
	\begin{subfigure}[t]{0.5\textwidth}
		\label{fig:parameters_alpha=1e-3_n=3_plot_barotropic}
		\flushright
		\includegraphics[width=215pt]{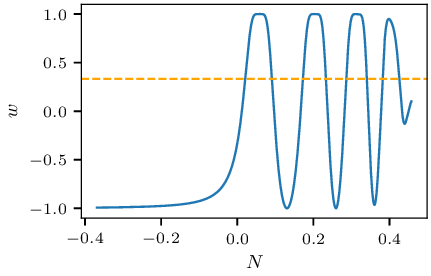}
	\end{subfigure} \\
	\begin{subfigure}[t]{0.5\textwidth}
		\label{fig:parameters_alpha=1e-3.5_n=2_plot_barotropic}
		\flushleft
		\includegraphics[width=215pt]{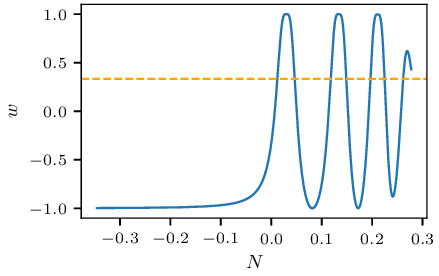}
	\end{subfigure}\hfill %
	\begin{subfigure}[t]{0.5\textwidth}
		\label{fig:parameters_alpha=1e-3.5_n=3_plot_barotropic}
		\flushright
		\includegraphics[width=215pt]{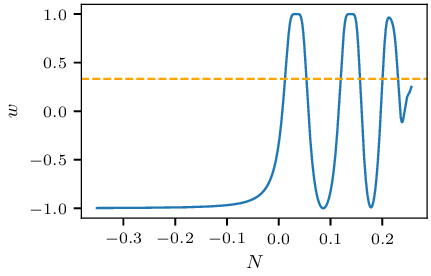}
	\end{subfigure} \\
	\begin{subfigure}[t]{0.5\textwidth}
		\label{fig:parameters_alpha=1e-4_n=2_plot_barotropic}
		\flushleft
		\includegraphics[width=215pt]{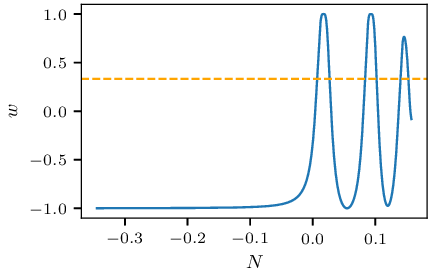}
	\end{subfigure}\hfill %
	\begin{subfigure}[t]{0.5\textwidth}
		\label{fig:parameters_alpha=1e-4_n=3_plot_barotropic}
		\flushright
		\includegraphics[width=215pt]{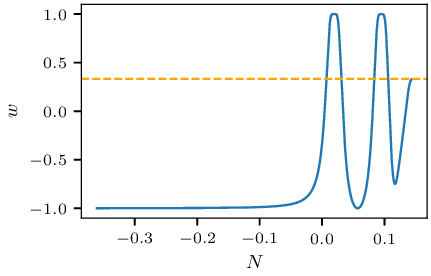}
	\end{subfigure} %
	\caption{Evolution of the barotropic parameter $w$ (blue solid line) as a function of the number of e-folds $N$ from the end of inflation ($\epsilon = 1$) for $n=2$ (left column) and $n=3$ (right column) and $\alpha = 10^{-3}$ (top), $\alpha=10^{-3.5}$ (middle), $\alpha=10^{-4}$ (bottom). Dashed orange vertical line correspond to $w=1/3$. \protect\label{fig:n=2,3_plot_barotropic}}
\end{figure*}

Another consequence of geometrical destabilization in models under consideration, which can be compared with analysis presented in \cite{Iarygina:2018kee}, are the spectra of excited fluctuations of the spectator field $\chi$. Obviously, the linear approximation used in \cite{Iarygina:2018kee} cannot capture the non-linear effect of rescattering of produced modes, which leads to broadening of the fluctuations spectra and production of inflaton fluctuations from those of the spectator. Hence, such a comparison should be made at the beginning of the geometrical destabilization.
We can see that the initial production of modes is compatible with the
results of the linear theory, and that the spectator spectrum displays a similar structure as presented in figure 14 of \cite{Iarygina:2018kee}. Both spectra are peaked for $k / \mu < 1$ as predicted by the linear theory, and geometrical destabilization initially affects only the modes with $k / \mu \lesssim 1$ for all considered models. 

\begin{figure*}[!ht]
	\begin{subfigure}[t]{0.5\textwidth}
		\label{fig:parameters_alpha=1e-3_n=2_plot_chi_fluct_N}
	    \flushleft
		\includegraphics[width=215pt]{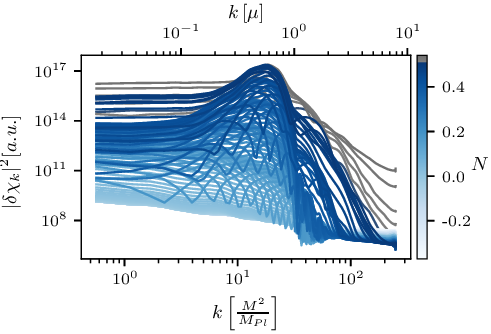}
	\end{subfigure} \hfill %
	\begin{subfigure}[t]{0.5\textwidth}
		\label{fig:parameters_alpha=1e-3_n=3_plot_chi_fluct_N}
		\flushright
		\includegraphics[width=215pt]{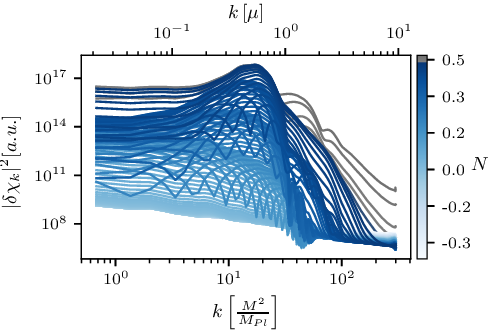}
	\end{subfigure} \\
	\begin{subfigure}[t]{0.5\textwidth}
		\label{fig:parameters_alpha=1e-3.5_n=2_plot_chi_fluct_N}
		\flushleft
		\includegraphics[width=215pt]{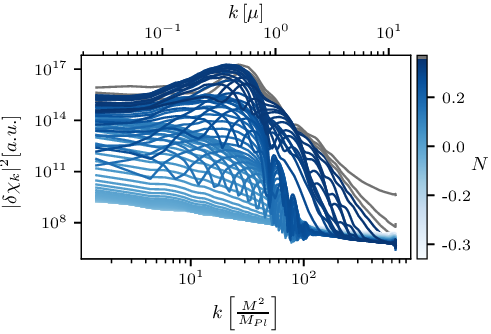}
	\end{subfigure}\hfill %
	\begin{subfigure}[t]{0.5\textwidth}
		\label{fig:parameters_alpha=1e-3.5_n=3_plot_chi_fluct_N}
		\flushright
		\includegraphics[width=215pt]{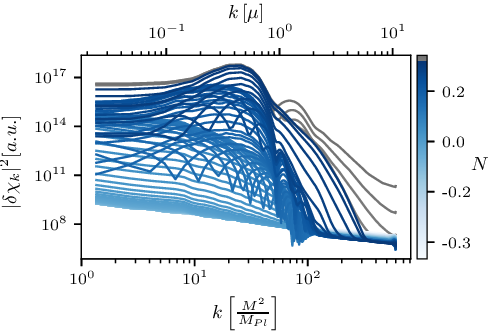}
	\end{subfigure}\\
	\begin{subfigure}[t]{0.5\textwidth}
		\label{fig:parameters_alpha=1e-4_n=2_plot_chi_fluct_N}
	    \flushleft
		\includegraphics[width=215pt]{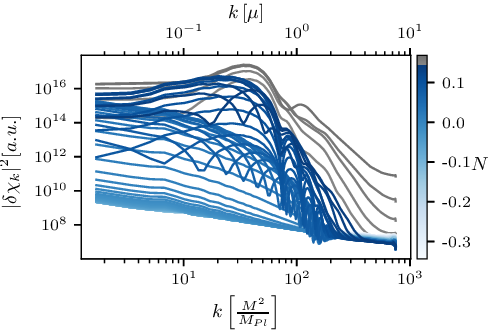}
	\end{subfigure}\hfill %
	\begin{subfigure}[t]{0.5\textwidth}
		\label{fig:parameters_alpha=1e-4_n=3_plot_chi_fluct_N}
		\flushright
		\includegraphics[width=215pt]{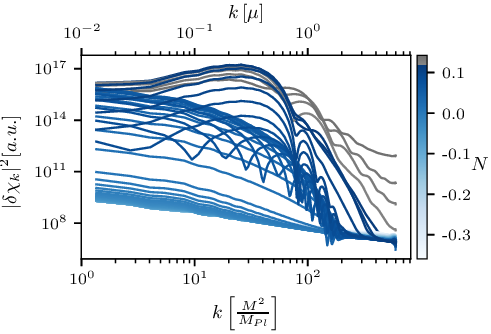}
	\end{subfigure} %
	\caption{Evolution of the spectrum of fluctuations $\delta \chi$ of the spectator field as a function of the number of e-folds $N$ from the end of inflation for $n=2$ (left column) and $n=3$ (right column) and $\alpha = 10^{-3}$ (top), $\alpha=10^{-3.5}$ (middle), $\alpha=10^{-4}$ (bottom). Grey lines denote
    results obtained after a potential breakdown of the simulations, as discussed in the text. Lines are drawn for time steps corresponding to 0.005 efolds. \protect\label{fig:n=2,3_plot_chi_fluct_N}}
\end{figure*}

\section{Spectrum of gravitational waves\label{sec:GWs_spectrum}}
\subsection{Numerical calculation of the spectrum}
\label{sec:s41}

The inhomogeneities of scalar fields produced during
preheating can act as a source of gravitational waves (GWs).
A numerical algorithm often used for computation of spectrum of GWs in numerical lattice simulations was proposed in \cite{Dufaux:2007pt}. According to  \cite{Dufaux:2007pt}, the spectrum can be estimated by solving the following ordinary differential equation in the Fourier space:
\begin{equation}
	{\partial_\tau}^2 \widehat{\bar{h}}_{ij} + \left(|k|^2- \frac{{\partial_\tau}^2 a}{a}\right) \widehat{\bar{h}}_{ij} = \frac{2 a}{{M_{Pl}}^2} \widehat{T^{TT}}_{ij}, \label{eq:transformed}
\end{equation}
where $\widehat{\bar{h}}_{ij}$ is a~Fourier transform of
\begin{equation}
	\bar{h} _{ij} = a h_{ij},
\end{equation}
which is a~transverse ($\partial_i h_{ij}=0$) and traceless ($h_{ii}=0$) part of the fluctuation around Friedman-Robertson-Walker metric:
\begin{equation}
	\mathrm{d}s^2 = \mathrm{d}t^2 - a^2(t) \left(\delta_{ij} + h_{ij}\right) \mathrm{d}x^i \mathrm{d}x^j = a^2(\tau) \left(\mathrm{d}\tau^2 -(\delta_{ij} + h_{ij}) \mathrm{d}x^i \mathrm{d}x^j\right)\, .
\end{equation}
The source term on the right hand side of eq. \eqref{eq:transformed} is the Fourier transform of a~transverse-traceless part $T^{TT}$ of the stress tensor $T$.
It can be expressed in a compact form as:
\begin{equation}
\widehat{T^{TT}}_{ij} (k) = \mathcal{O}_{ijlm} (k) \widehat{T}_{ij} (k)\,, \label{tranverse-traceless}
\end{equation}
using the projection operators
\begin{subequations}\label{projection_O}
\begin{equation}
\mathcal{O}_{ijlm} (k) = P_{il} (k) P_{jm} (k) - \frac{1}{2} P_{ij} (k) P_{lm} (k)\,,
\end{equation}
where
\begin{equation}
P_{ij}(k) = \delta_{ij} - \frac{k_i k_j}{|k|^2}\,.
\end{equation}
\end{subequations}
In models described by the Lagrangian \eqref{eq:field_space_metric}, such
as T-models of $\alpha$-attractors,
the lowest order contribution to $T^{TT}$ in expansion in powers of $h$ is given by:
\begin{equation}
    T_{ij} = e^{2b(\chi)} \partial_{i} \phi \partial_{j} \phi + \partial_{i} \chi \partial_{j} \chi + \BigO{h} + \text{homogeneous terms}
\end{equation}
According to \cite{Dufaux:2007pt}, the energy density of GWs per mode is given by:
\begin{equation}
	\varrho_{GW}(\tau_f,k) = \frac{{M_{Pl}}^2}{4 \pi {a(\tau_f)}^4 V} \sum_{i,j} \left(\left|\left(\partial_\tau \widehat{\bar{h}}_{ij}\right)(\tau_f,k)\right|^2 + |k|^2 \left|\widehat{\bar{h}}_{ij}(\tau_f,k)\right|^2\right),
\end{equation}
where $\tau_f$ is a conformal time at which the source disappears. 

In this paper, we are mostly interested in the spectrum of the energy density of GWs per unit logarithmic frequency interval:
\begin{equation}
\frac{d \rho_{GW}}{d \log |k|} (\tau ,k) = |k|^3 \int_{S^2} d \Omega \varrho_{gw}(\tau, k), \label{energy_density}
\end{equation}
where $\int_{S^2} d \Omega$ denotes the integration over the direction of the wave vector $k$.

In \cite{Dufaux:2007pt},
the time evolution of the scale factor $a$ was approximated as proportional
to the conformal time, 
as during radiation domination.
In such a case, the eq.~\eqref{eq:transformed} can be solved using the retarded Green's function:
\begin{equation}
\widehat{\bar{h}}_{ij}(\tau,k) = \frac{2}{{M_{Pl}}^2} \int_{\tau_i}^{\tau} d\tau' \frac{\sin \left(|k|\left(\tau - \tau'\right)\right)}{|k|} a(\tau') \widehat{T^{TT}}_{ij} (\tau',k),\label{source_solution}
\end{equation}
where $\tau_i$ is the value of conformal time before which the source $\widehat{T^{TT}}_{ij}$ appeared. 
However, in our numerical simulations, the  scale factor $a$ is a~dynamical variable, whose evolution deviates from simple linear dependence $a \propto \tau$. In order to take into account this effect, we developed an algorithm for direct integration of the eq. \eqref{eq:transformed}. We present it in the appendix~\ref{app:sec:numerical_scheme_for_GWs}. Similar attempts were done in the past \cite{Easther:2006vd, Garcia-Bellido:2007fiu, Easther:2007vj} using general purpose Runge-Kutta integrator.

The spectrum of GWs calculated from a~lattice simulation needs to be properly redshifted in order to obtain the spectrum at the present time for comparison with sensitivities of GWs detectors and exclusion limits.
According to \cite{Dufaux:2007pt}, a monochromatic GW with comoving wave-vector $k$ will correspond to wave with the present-day frequency 
\begin{equation}
    f = {a_f}^{-1} \left( \frac{k}{\rho_f^{1/4}} \right) \left( \frac{a_f}{a_*} \right)^{1 - \frac{3}{4}(1+w)} 4 \times 10^{10}\, \textrm{Hz}, \label{eq:GW_frequency_reshift}
\end{equation}
where $a_f$, $\rho_f$ are values of, respectively, the scale factor and the critical energy density at the time $\tau_f$ of end of the simulation and  $a_*$ is the value of scale factor at time $\tau_*$ when thermal equilibrium is established. This expression was derived under the assumption that the Universe evolved from $\tau_f$ to $\tau_*$ with a mean effective barotropic parameter equal to $w$. As we showed in Section \ref{sec:results_of_numerical_simulations}, we can expect that the reheating in most $\alpha$-attractor T-models with considered parameters is instantaneous and that the effective barotropic parameter reaches value $\frac{1}{3}$ within one e-fold from the end of the inflation. Thus, we shall presumably make a negligible error, ignoring the factor $\frac{a_f}{a_*}$ in \eqref{eq:GW_frequency_reshift}. The energy density spectrum today is given by
\begin{equation}
    \Omega_{GW} h^2 (k) = 9.3 \times 10^{-6}\rho_f^{-1} \frac{d \rho_{GW}}{d \log |k|} (\tau_f ,k).
\end{equation}

Frequencies \eqref{eq:GW_frequency_reshift} of GWs produced during preheating are generically predicted to be very large in comparison with sensitivity ranges of ground-based interferometric detectors (LIGO/Virgo \cite{Thrane:2013oya, TheLIGOScientific:2014jea, TheLIGOScientific:2016wyq, LIGOScientific:2019vic} or ET \cite{Punturo:2010zz, Hild:2010id}) for GWs generated around 
inflation.
Although certain ways of detecting GWs at very high frequencies were proposed~\cite{Aggarwal:2020olq}, their predicted reach in terms of abundance is still above the current lower bound on the amplitude of GWs coming from estimation of number of relativistic degrees of freedom $N_\text{eff}$ during Big Bang nucleosynthesis (BBN) based on CMB observations \cite{Smith:2006nka, Henrot-Versille:2014jua}. Current constraint from the Planck 2018 results, $N_\text{eff}=3.27 \pm 0.15$ \cite{Planck:2018vyg}, requires that $\Omega_{GW} h^2 \le 1.85 \times 10^{-6}$ and we use this bound in the plots of spectra of GWs in the remainder of our analysis.

\subsection{Results from numerical calculations}
We have computed the spectra of gravitational waves emitted during preheating in lattice simulations using a numerical algorithm described in Section \ref{sec:GWs_spectrum}. We directly integrated in time equation \eqref{eq:transformed}, using the integrator based on Magnus expansion presented in the appendix \ref{app:sec:numerical_scheme_for_GWs}. 
Our results for $\alpha=10^{-3}, 10^{-3.5}, 10^{-4}$ are presented in figure \ref{fig:n=1,1.5_plot_spectrum} for $n=1, 1.5$ and in figure \ref{fig:n=2,3_plot_spectrum} for $n=2, 3$. 
For more informative character, we plotted the spectra in various moments of time, i.e. we show how the spectrum of GWs would look like today if its production out of scalar fluctuations stopped at given time. Similarly as for spectra of fluctuations of fields $\phi$ and $\chi$, the color of the line encodes the time. The red band visible at the top of some figures correspond to BBN exclusion limit discussed in Section \ref{sec:s41}.

\begin{figure*}[!ht]
	\begin{subfigure}[t]{0.5\textwidth}
		\label{fig:parameters_alpha=1e-3_n=1_plot_spectrum_N}
	    \flushleft
		\includegraphics[width=215pt]{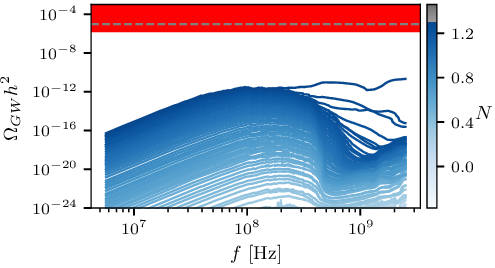}
	\end{subfigure} \hfill %
	\begin{subfigure}[t]{0.5\textwidth}
		\label{fig:parameters_alpha=1e-3_n=1.5_plot_spectrum_N}
		\flushright
		\includegraphics[width=215pt]{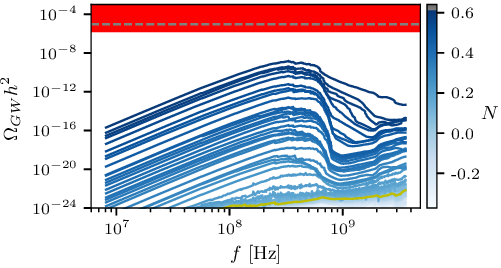}
	\end{subfigure} \\
	\begin{subfigure}[t]{0.5\textwidth}
		\label{fig:parameters_alpha=1e-3.5_n=1_plot_spectrum_N}
		\flushleft
		\includegraphics[width=215pt]{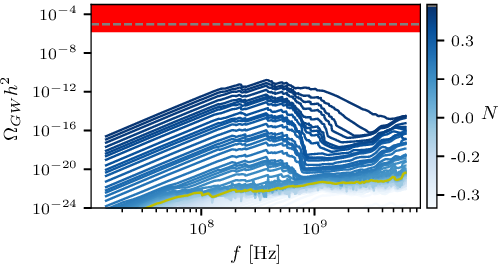}
	\end{subfigure}\hfill %
	\begin{subfigure}[t]{0.5\textwidth}
		\label{fig:parameters_alpha=1e-3.5_n=1.5_plot_spectrum_N}
		\flushright
		\includegraphics[width=215pt]{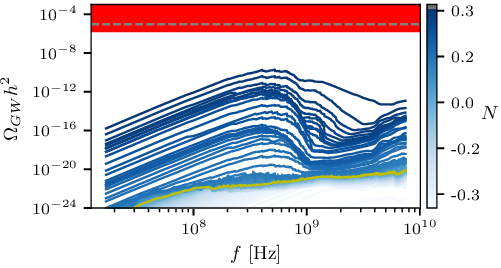}
	\end{subfigure}\\
	\begin{subfigure}[t]{0.5\textwidth}
		\label{fig:parameters_alpha=1e-4_n=1_plot_spectrum_N}
	    \flushleft
		\includegraphics[width=215pt]{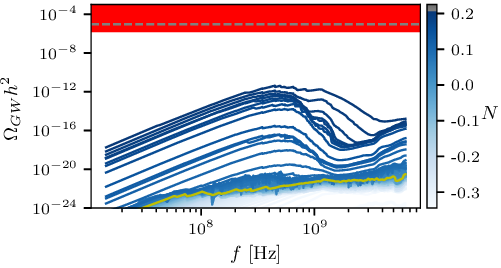}
	\end{subfigure}\hfill %
	\begin{subfigure}[t]{0.5\textwidth}
		\label{fig:parameters_alpha=1e-4_n=1.5_plot_spectrum_N}
		\flushright
		\includegraphics[width=215pt]{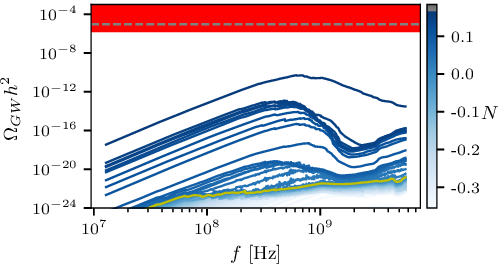}
	\end{subfigure} %
	\caption{Evolution of the spectrum of gravitational waves as a function of the number of e-folds $N$ from the end of inflation for $n=1$ (left column) and $n=1.5$ (right column) and $\alpha = 10^{-3}$ (top), $\alpha=10^{-3.5}$ (middle), $\alpha=10^{-4}$ (bottom).
		Lines of different colors correspond to different time points
	described by the number of e-folds after the end of inflation,
	as indicated on the bar next to each panel. Yellow lines correspond
    to the end of inflation at $N=0$.  Lines are drawn for time steps corresponding to 0.005 efolds.
	The red bands correspond to BBN exclusion limit and horizontal dashed grey lines correspond to GWs dominating the energy density of the Universe, as discussed in Section \ref{sec:GWs_spectrum}.
	\protect\label{fig:n=1,1.5_plot_spectrum}}
\end{figure*}

\begin{figure*}[!ht]
	\begin{subfigure}[t]{0.5\textwidth}
		\label{fig:parameters_alpha=1e-3_n=2_plot_spectrum_N}
		\flushleft
		\includegraphics[width=215pt]{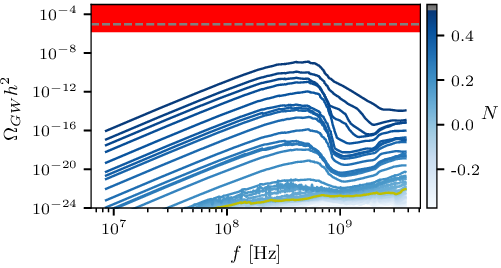}
	\end{subfigure}\hfill %
	\begin{subfigure}[t]{0.5\textwidth}
		\label{fig:parameters_alpha=1e-3_n=3_plot_spectrum_N}
		\flushright
		\includegraphics[width=215pt]{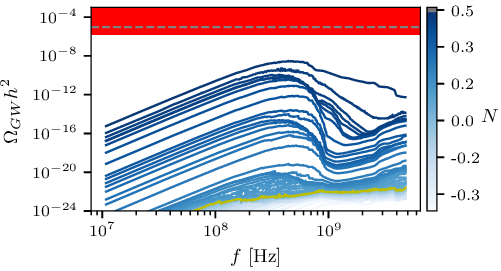}
	\end{subfigure} \\
	\begin{subfigure}[t]{0.5\textwidth}
		\label{fig:parameters_alpha=1e-3.5_n=2_plot_spectrum_N}
		\flushleft
		\includegraphics[width=215pt]{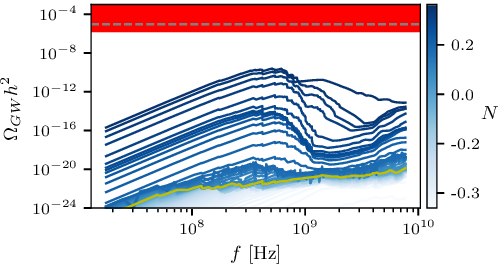}
	\end{subfigure}\hfill %
	\begin{subfigure}[t]{0.5\textwidth}
		\label{fig:parameters_alpha=1e-3.5_n=3_plot_spectrum_N}
		\flushright
		\includegraphics[width=215pt]{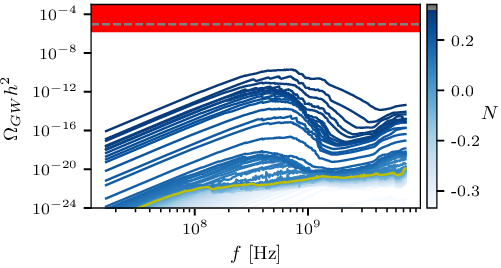}
	\end{subfigure} \\
	\begin{subfigure}[t]{0.5\textwidth}
		\label{fig:parameters_alpha=1e-4_n=2_plot_spectrum_N}
		\flushleft
		\includegraphics[width=215pt]{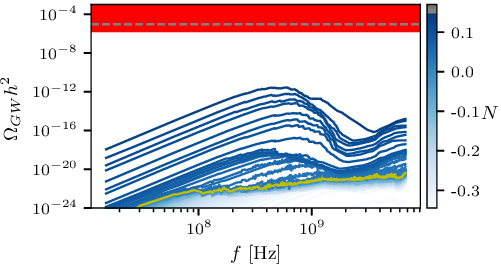}
	\end{subfigure}\hfill %
	\begin{subfigure}[t]{0.5\textwidth}
		\label{fig:parameters_alpha=1e-4_n=3_plot_spectrum_N}
		\flushright
		\includegraphics[width=215pt]{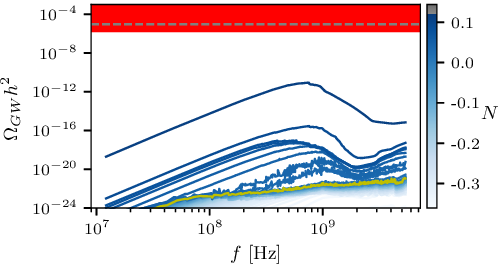}
	\end{subfigure} %
	\caption{Evolution of the spectrum of gravitational waves as a function of the number of e-folds $N$ from the end of inflation for $n=2$ (left column) and $n=3$ (right column) and $\alpha = 10^{-3}$ (top), $\alpha=10^{-3.5}$ (middle), $\alpha=10^{-4}$ (bottom).
		Lines of different colors correspond to different time points
	described by the number of e-folds after the end of inflation,
	as indicated on the bar next to each panel. Yellow lines correspond
    to the end of inflation at $N=0$.  Lines are drawn for time steps corresponding to 0.005 efolds.
	The red band visible at the top of some figures correspond to BBN exclusion limit and horizontal dashed grey line to the case when GWs dominate the energy density as discussed in section \ref{sec:GWs_spectrum}.
	\protect\label{fig:n=2,3_plot_spectrum}}
\end{figure*}

As we showed in Section \ref{sec:results_of_numerical_simulations}, the process of reheating in two-fields $\alpha$-attractor T-models has a certain pattern dictated by the geometrical destabilization of the spectator field. At the end of inflation when $\epsilon \sim 1$ the negative curvature of the field space produce the negative contribution to the effective mass of the spectator, driving this mass tachyonic and leading to the production of spacial fluctuations of this field. During reheating the energy density of the homogeneous inflaton background is transferred into excitations of the spectator. Finally, when the significant fraction of the energy is siphoned off by the spectator, the nonlinear coupling of fields leads to production of inflaton fluctuations and rescattering of spectator ones into higher momenta modes. This pattern has its implications for production of GWs from fluctuations of fields.

As is clearly visible from plots in figures \ref{fig:n=1,1.5_plot_spectrum} and \ref{fig:n=2,3_plot_spectrum}, the spectra computed for times before the end of inflation ($N<0$) saturate at extremely low level of $10^{-24}$---$10^{-20}$ which probably correspond to the numerical precision with which we reproduce the amplitudes of Bunch--Davis vacuum fluctuations.

After the end of inflation ($N=0$), corresponding to the yellow
lines in figures \ref{fig:n=1,1.5_plot_spectrum} and \ref{fig:n=2,3_plot_spectrum}, the spectra start to grow and develop well established maxima which are located at the wave vectors $k$ of the order of twice the location of maxima of spectra of fluctuations of the spectator field, since the two scalar modes are needed for production of GW. The amplitudes of the spectra at this stage are roughly of the order of $10^{-10}$--$10^{-9}$, typical for the resonance- and tachyonic instability-driven preheating in inflation at these scales \cite{Amin:2018kkg, Adshead:2018doq, Adshead:2019igv, Adshead:2019lbr, Sang:2019ndv, Lozanov:2019ylm, Li:2020qnk, Ghoshal:2022jdt} (further references to works studying production of GWs during preheating can be found in section 3.3.2 of \cite{Aggarwal:2020olq}).

For times around the end of the simulation, 
high-frequency GWs are massively produced.
This rapid growth of high energy GWs is associated with production of 
high-$k$ modes of the inflaton field due to nonlinear interaction 
of spectator and inflaton fluctuations.
However, as discussed in section 
\ref{sec:results_of_numerical_simulations},
the fast-developing UV cascade quickly excites
even the shortest wavelengths that we can resolve in our simulations,
which potentially signals a breakdown of the simulations.
Therefore, we 
present only the results obtained before such a breakdown occurs,
determined with
the criterion introduced in section~\ref{sec:results_of_numerical_simulations}.
We also note that it was pointed out that 
the underlying instability of the scalar fluctuations cannot be fully 
resolved in fixed-lattice simulations \cite{Li:2020qnk}. 

In order to ensure maximal reliability of our results, we compared 
spectra computed in simulations with different cutoffs, 
and found no qualitative differences. 
Therefore, we believe that we can treat our results as trustworthy, especially
that we assumed two conservative conditions 
for stopping our simulations. 
This comes with a price of possibly underestimating the GW spectra, 
since the production of GWs will proceed longer
than the time span of our simulation.
With all these caveats in mind, we conclude that the
amount of gravitational waves produced in the 
two-field $\alpha$-attractor T-models
that we analyzed peaks at $\Omega_{\rm GW}h^2\sim 10^{-12}-10^{-9}$.
These values are three order of magnitude below the BBN bound
on the presence of additional relativistic degrees of freedom.

We note that the uppermost part of figures \ref{fig:n=1,1.5_plot_spectrum} and \ref{fig:n=2,3_plot_spectrum}, above $\Omega_{\rm GW}>10^{-5}$,
do not correspond to a self-consistent calculation, because
such quantities of GWs today would correspond
to GWs domination the energy density of the Universe
at preheating. This is of no particular importance for our results,
since we do not claim to have obtained reliable
predictions for GWs in that ballpark.

\section{Conclusions and outlook}

In this paper, we extended our earlier investigation of (p)reheating in
two-field $\alpha$-attractor models, reported in \cite{Krajewski:2018moi},
in two directions.
We extended our lattice simulations to a larger $512^3$ lattice, refining
certain technical aspects to make sure that our results are reliable.
We corroborated the fact that fluctuations of the spectator
field are highly unstable after inflation and quickly grow,
with their energy density catching up with that of the homogeneous
inflaton. These fluctuations enter nonlinear regime and become
a driving force for the growth of inflaton fluctuations. 
We also find that the growth of the fluctuations is so large and encompasses such a wide range of modes that it becomes
a formidable computational challenge to resolve this nonlinear dynamics
on the lattice.

We also calculated the power spectra of gravitational waves produced
out of the fluctuations of the scalar fields. We found that
large amplitudes of the scalar fields translate into
large amplitudes of gravitational waves at frequencies above
the range that is probed or will soon be
with `direct detection' experiments. 
The strength of the instability of the scalar fluctuations
proved to be a formidable challenge to track the evolution
of these fluctuations. 
From the conservative point of view, the present-day 
amount of GWs
is three orders of magnitude smaller than the upper
constraints on the number of the extra
relativistic degrees of freedom, inferred from the CMB and BBN data,
if we evaluate the scalar field fluctuations
at the instant when our simulations appear to break down.

The research reported in this paper is just a step towards
a fuller understanding of two-field $\alpha$-attractor T-models
and can be extended in several
 directions. In spite of a significant increase of
the lattice size and the use of a powerful computational facility,
we have been unable to fully resolve the instability. This suggests
that it is necessary to go beyond the fixed lattice calculations and
to perform multi-scale analysis of the instability. Such a new numerical
tool would allow giving more accurate predictions about the dynamics
of the fluctuations of both the inflaton and the spectator, and
about the effective equation of state of the Universe during
post-inflationary geometrical destabilization. It would also enable
a more refined scan over the parameter space, greatly
reducing theoretical uncertainty of the inflationary
predictions for the $\alpha$-attractor T-models. 
In particular, a better resolution of the instability 
would offer better predictions for the amplitudes of GWs,
going beyond the lower bounds presented in this paper and
addressing the question whether 
the energy density released in the form 
of GWs can be incompatible with the BBN bound, ruling out
some $\alpha$-attractor T-models. 
A possibility of producing primordial black holes out of
exponentially growing curvature perturbations could also
be addressed quantitatively. Last but not least, coupling of scalar fields to full gravity (as was done for a different inflationary model in \cite{Giblin:2019nuv}) may be needed at late time when energy density of spacial fluctuations of fields dominate Universe, since backreaction from fluctuations of metric tensor probably influence the process as was speculated in \cite{Joana:2022uwc}. Albeit interesting, all these
goals require a novel numerical approach and construction 
of beyond state-of-the-art numerical tools. We shall therefore
consider them a challenge to be addressed in future research.

\acknowledgments{
T.K.~is supported by grant 2019/32/C/ST2/00248 from the National Science Centre (Poland).
K.T.~is partially supported by grant 2018/30/Q/ST9/00795 from the National Science Centre (NCN).
This research was supported in part by PL--Grid Infrastructure. We would like to thank Jan Henryk Kwapisz for his kind help in implementation of some numerical tools used during preparation of this manuscript.
}

\appendix

\section{Numerical scheme for calculation of spectra of gravitational waves in non-trivial background metric evolution\label{app:sec:numerical_scheme_for_GWs}}

\subsection{Construction of the integrator}
The method presented here is implementation of the second order scheme based on Magnus expansion presented in \cite{Blanes:2012}.

Eq.~\eqref{eq:transformed} can be rewritten as a~first order non-homogeneous ordinary differential equation:
\begin{equation}
\begin{split}
	{\partial_\tau} Y :=& \,\,\partial_\tau
	\begin{bmatrix}
		\widehat{\bar{h}}_{ij} \\
		\partial_\eta \widehat{\bar{h}}_{ij}
	\end{bmatrix} =
	\begin{bmatrix}
		0 & 1 \\
		-|k|^2 + \frac{{\partial_\tau}^2 a}{a} & 0
	\end{bmatrix}
	\begin{bmatrix}
		\widehat{\bar{h}}_{ij} \\
		\partial_\eta \widehat{\bar{h}}_{ij}
	\end{bmatrix}
	+
	\begin{bmatrix}
		0 \\
		\frac{2 a}{{M_{Pl}}^2} \widehat{T^{TT}}_{ij}
	\end{bmatrix} \\
	=& \,\,M Y + F. \label{eq:first_order}
	\end{split}
\end{equation}
According to \cite{Blanes:2012}, a formal solution of \eqref{eq:first_order} is of the form:
\begin{equation}
	Y(\tau + \delta \tau) = \Phi(\tau + \delta \tau, \tau) Y(\tau) + \int_{\tau}^{\tau + \delta \tau} \Phi(\tau + \delta \tau, \xi) F(\xi) d\xi\,, \label{eq:formal}
\end{equation}
where $\Phi$ is the fundamental solution matrix of the associated homogeneous equation:
\begin{equation}
	\partial_\tau \Phi(\tau, \tau_0) = M(\tau) \Phi(\tau, \tau_0). \label{eq:fundamental}
\end{equation}
Using Magnus approximation in the lowest order, the fundamental solution of eq.~\eqref{eq:fundamental} can be represented as:
\begin{equation}
	\Phi(\tau + \delta \tau, \eta) = \exp{\left( \int_{\tau}^{\tau + \delta \tau} M(\xi) d\xi + \BigO{\delta \tau^4}\right)}. \label{eq:Magnus_approximation}
\end{equation}
One can use the trapezoid rule to approximate integrals in eqs.~\eqref{eq:Magnus_approximation} and \eqref{eq:formal} in order to obtain an explicit, time-reversible, second-order discretization scheme:
\begin{equation}
	Y_{n + 1} = \exp{\left( \frac{\delta \tau}{2} (M_{n + 1} + M_{n})\right)} \left( Y_{n} + \frac{\delta \tau}{2} F_{n} \right) + \frac{\delta \tau}{2} F_{n + 1} \label{eq:scheme}
\end{equation}
given in \cite{Blanes:2012}. Applying \eqref{eq:scheme} to the equation of motion of metric perturbations \eqref{eq:first_order} gives  the following set of equations:
\begin{equation}
	\begin{split}
		\widehat{\bar{h}}_{ij, n + 1} =& \cos{(\omega \delta \tau)} \widehat{\bar{h}}_{ij, n} + \frac{i}{\omega} \sin{(\omega \delta \tau)} \left(\partial_\tau \widehat{\bar{h}}_{ij, n} + \delta \tau \frac{a_{n}}{{M_{Pl}}^2} \widehat{T^{TT}}_{ij, n} \right) \\
		\partial_\tau \widehat{\bar{h}}_{ij, n + 1} =& \cos{(\omega \delta \tau)} \left(\partial_\tau \widehat{\bar{h}}_{ij, n} + \delta \tau \frac{a_{n}}{{M_{Pl}}^2} \widehat{T^{TT}}_{ij, n} \right) + i \omega \sin{(\omega \delta \tau)} \widehat{\bar{h}}_{ij, n}\\
		&+\delta \tau \frac{a_{n + 1}}{{M_{Pl}}^2} \widehat{T^{TT}}_{ij, n + 1}\,, \label{eq:dicrete}
	\end{split}
\end{equation}
where the frequency $\omega$ is defined as:
\begin{equation}
	\omega = \sqrt{|k|^2 - \frac{{\partial_\tau}^2 a_{n + 1}}{2a_{n + 1}} - \frac{{\partial_\tau}^2 a_{n}}{2a_{n}}}\,,
\end{equation}
and $n$ numerates time steps.

\subsection{Error estimation for the integrator}
\label{app:error}
In order to use the theory of error estimation for Magnus expansion integrator developed in \cite{Iserles:1999} one need to rewrite eq.~\eqref{eq:first_order} as a~homogeneous equation as was described in \cite{Blanes:2012}:
\begin{equation}
	{\partial_\tau} Z := \partial_\tau
	\begin{bmatrix}
		Y \\
		1
	\end{bmatrix} =
	\begin{bmatrix}
		M & F \\
		0 & 0
	\end{bmatrix}
	\begin{bmatrix}
		Y \\
	1
	\end{bmatrix}
	= A Z. \label{eq:homogeneous}
\end{equation}
The errors in the integration with discretization scheme given by the eq.~\eqref{eq:dicrete} come from two sources: from the truncation of the Magnus expansion and from the trapezoid rule.

We use the error introduced by the first source to adaptatively choose integration time step, since the estimation of the second requires computation of time derivatives of the stress tensor. The next term in the Magnus expansion which we neglected constructing second order scheme is of the form:
\begin{equation}
	\Omega^{[4]} - \Omega^{[2]} = \left[A^{(1)}, A^{(0)}\right],
\end{equation}
where $[\cdot, \cdot]$ is a commutator of matrices and $A^{(i)}$ is defined as:
\begin{equation}
	A^{(i)} := \frac{1}{\delta \tau^i} \int_\tau^{\tau + \delta \tau} \left(\xi - \tau - \frac{\delta \tau}{2}\right)^i A(\xi) d\xi.
\end{equation}
In our case, the commutator gives:
\begin{equation}
	\Omega^{[4]} - \Omega^{[2]} =
	\begin{bmatrix}
		- h \left(\frac{{\partial_\tau}^2 a}{a}\right)^{(1)} & 0 & -h\left(\frac{2 a}{{M_{Pl}}^2} \widehat{T^{TT}}_{ij}\right)^{(1)} \\
		0 & h \left(\frac{{\partial_\tau}^2 a}{a}\right)^{(1)} & 0 \\
		0 & 0 & 0
	\end{bmatrix}.
\end{equation}
We can use the trapezoidal rule to obtain the lowest order in $\delta \tau$ approximation to the integrals in the error estimation:
\begin{align}
	\left(\frac{{\partial_\tau}^2 a}{a}\right)^{(1)} &= \frac{\delta \tau}{4} \left(\frac{{\partial_\tau}^2 a_{n + 1}}{a^{n + 1}} - \frac{{\partial_\tau}^2 a_{n}}{a_{n}}\right) + \BigO{\delta \tau^2}, \\
	\left(\frac{2 a}{{M_{Pl}}^2} \widehat{T^{TT}}_{ij}\right)^{(1)} &= \frac{\delta \tau}{2 {M_{Pl}}^2} \left(\widehat{T^{TT}}_{ij, n + 1} - \widehat{T^{TT}}_{ij, n}\right) + \BigO{\delta \tau^2}
\end{align}
Finally, we need to translate the error estimation for the exponent $\Omega$ into expressions for errors of the components of $Y$ vector. Let us denote $Z(\tau +\delta \tau) = \exp{(\Omega(\tau + \delta \tau))} Z_0$ as an~exact solution to eq.~\eqref{eq:first_order} and $\widetilde{Z}_n = \exp{(\widetilde{\Omega}_n)} Z_0$ as our approximation calculated using discretization \eqref{eq:dicrete}. Expressing initial $Z_0$ by $\widetilde{Z}_n$ with the inverse of the evolution operator, one gets:
\begin{equation}
	Z(\tau + \delta \tau) - \widetilde{Z}_{n + 1} = \left(\exp{(\Omega(\tau + \delta \tau))} \exp{(-\widetilde{\Omega}_n)} - 1\right) \widetilde{Z}_{n + 1}.
\end{equation}
An expansion based on Baker-Campbell-Hausdorff formula leads to:
\begin{equation}
Z(\tau + \delta \tau) - \widetilde{Z}_{n + 1} = \left(\exp{\left(\Omega^{[4]}_n - \Omega^{[2]}_n \right)} - 1\right) \widetilde{Z}_{n + 1} + \BigO{\delta \tau^4}.
\end{equation}
Due to small dimension of $Z$ the exponent $\exp{\left(\Omega^{[4]}_n - \Omega^{[2]}_n\right)}$ can be calculated exactly giving explicit expressions for error estimates, as follows:
\begin{align}
	\delta \widehat{\bar{h}}_{ij, n + 1} =& \left(\exp{\left(-\frac{\delta \tau^2}{4} \left(\frac{{\partial_\tau}^2 a_{n + 1}}{a_{n + 1}} - \frac{{\partial_\tau}^2 a_{n}}{a_{n}}\right)\right)} - 1\right) \widehat{\bar{h}}_{ij_, n + 1} \\
	 &+ \frac{\frac{2}{{M_{Pl}}^2} \left(\widehat{T^{TT}}_{ij, n + 1} - \widehat{T^{TT}}_{ij, n}\right)}{\left(\frac{{\partial_\tau}^2 a_{n + 1}}{a_{n + 1}} - \frac{{\partial_\tau}^2 a_{n}}{a_{n}}\right)} \\
	 &\left(\exp{\left(-\frac{\delta \tau^2}{4} \left(\frac{{\partial_\tau}^2 a_{n + 1}}{a_{n + 1}} - \frac{{\partial_\tau}^2 a_{n}}{a_{n}}\right)\right)} - 1\right) \\
	 \delta \partial_\tau \widehat{\bar{h}}_{ij, n + 1} =& \left(\exp{\left(\frac{\delta \tau^2}{4} \left(\frac{{\partial_\tau}^2 a_{n + 1}}{a_{n + 1}} - \frac{{\partial_\tau}^2 a_{n}}{a_{n}}\right)\right)} - 1\right) \partial_\tau \widehat{\bar{h}}_{ij, n + 1}
\end{align}

\section{Supplementary results}
In this section, we provide further results from simulations with $n=2$ and $n=3$ which support our hypothesis, but are qualitatively analogous to cases $n=1$ and $n=1.5$ discussed in Section \ref{sec:results_of_numerical_simulations}. We believe that they can be useful for other researchers working in the field.

\begin{figure*}[!ht]
	\begin{subfigure}[t]{0.5\textwidth}
		\label{fig:parameters_alpha=1e-3_n=2_plot_energy}
	    \flushleft
		\includegraphics[width=215pt]{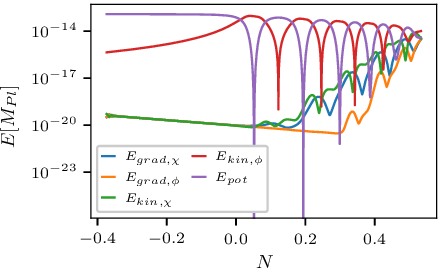}
	\end{subfigure} \hfill %
	\begin{subfigure}[t]{0.5\textwidth}
		\label{fig:parameters_alpha=1e-3_n=3_plot_energy}
		\flushright
		\includegraphics[width=215pt]{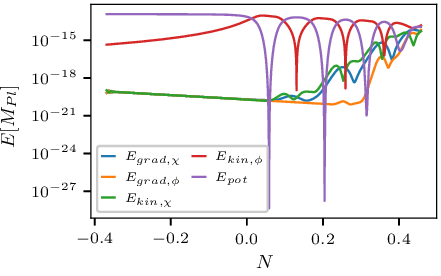}
	\end{subfigure} \\
	\begin{subfigure}[t]{0.5\textwidth}
		\label{fig:parameters_alpha=1e-3.5_n=2_plot_energy}
		\flushleft
		\includegraphics[width=215pt]{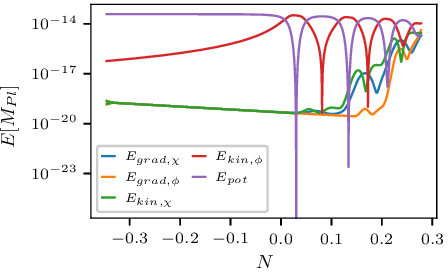}
	\end{subfigure}\hfill %
	\begin{subfigure}[t]{0.5\textwidth}
		\label{fig:parameters_alpha=1e-3.5_n=3_plot_energy}
		\flushright
		\includegraphics[width=215pt]{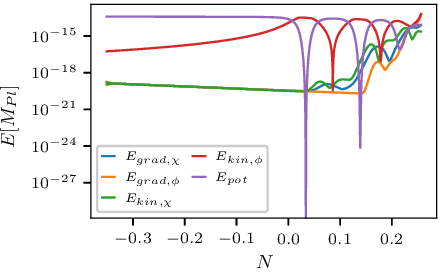}
	\end{subfigure}\\
	\begin{subfigure}[t]{0.5\textwidth}
		\label{fig:parameters_alpha=1e-4_n=2_plot_energy}
	    \flushleft
		\includegraphics[width=215pt]{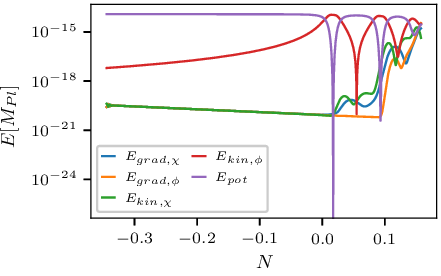}
	\end{subfigure}\hfill %
	\begin{subfigure}[t]{0.5\textwidth}
		\label{fig:parameters_alpha=1e-4_n=3_plot_energy}
		\flushright
		\includegraphics[width=215pt]{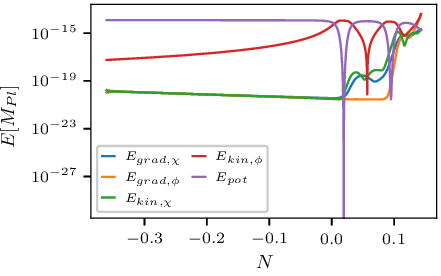}
	\end{subfigure} %
	\caption{Evolution of energy components as a function of the number of e-folds $N$ from the end of inflation for $n=2$ (left column) and $n=3$ (right column) and $\alpha = 10^{-3}$ (top), $\alpha=10^{-3.5}$ (middle), $\alpha=10^{-4}$ (bottom). \protect\label{fig:n=2,3_plot_energy}}
\end{figure*}

\begin{figure*}[!ht]
	\begin{subfigure}[t]{0.5\textwidth}
		\label{fig:parameters_alpha=1e-3_n=2_plot_phi_fluct_N}
	    \flushleft
		\includegraphics[width=215pt]{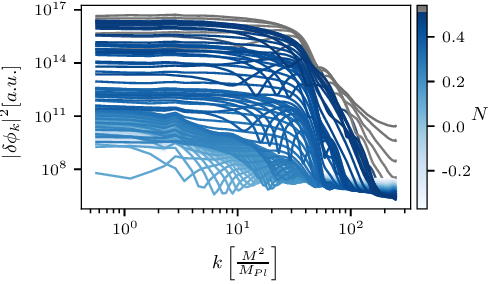}
	\end{subfigure} \hfill %
	\begin{subfigure}[t]{0.5\textwidth}
		\label{fig:parameters_alpha=1e-3_n=3_plot_phi_fluct_N}
		\flushright
		\includegraphics[width=215pt]{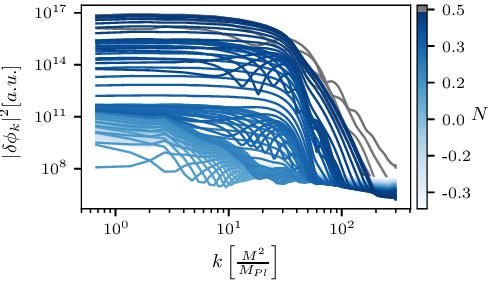}
	\end{subfigure} \\
	\begin{subfigure}[t]{0.5\textwidth}
		\label{fig:parameters_alpha=1e-3.5_n=2_plot_phi_fluct_N}
		\flushleft
		\includegraphics[width=215pt]{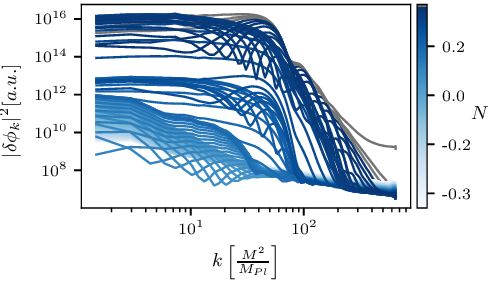}
	\end{subfigure}\hfill %
	\begin{subfigure}[t]{0.5\textwidth}
		\label{fig:parameters_alpha=1e-3.5_n=3_plot_phi_fluct_N}
		\flushright
		\includegraphics[width=215pt]{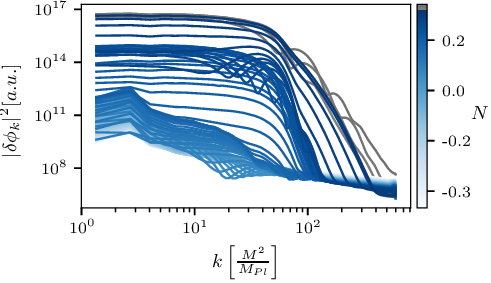}
	\end{subfigure}\\
	\begin{subfigure}[t]{0.5\textwidth}
		\label{fig:parameters_alpha=1e-4_n=2_plot_phi_fluct_N}
	    \flushleft
		\includegraphics[width=215pt]{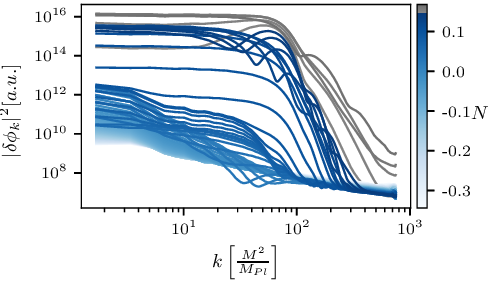}
	\end{subfigure}\hfill %
	\begin{subfigure}[t]{0.5\textwidth}
		\label{fig:parameters_alpha=1e-4_n=3_plot_phi_fluct_N}
		\flushright
		\includegraphics[width=215pt]{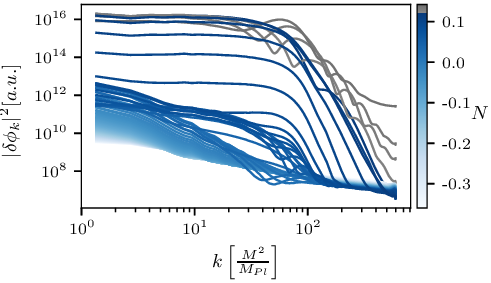}
	\end{subfigure} %
	\caption{Evolution of the spectrum of fluctuations $\delta \phi$ of the spectator field as a function of the number of e-folds $N$ from the end of inflation for $n=2$ (left column) and $n=3$ (right column) and $\alpha = 10^{-3}$ (top), $\alpha=10^{-3.5}$ (middle), $\alpha=10^{-4}$ (bottom). \protect\label{fig:n=2,3_plot_phi_fluct_N}}
\end{figure*}

\bibliographystyle{JHEP}
\bibliography{references}
\end{document}